# Metasurface-Enabled Multifunctional Single-Frequency Sensors without External Power

*Masaya Tashiro[1], Kosuke Ide[1], Kosei Asano[1], Satoshi Ishii[2, 3, 4], Yuta Sugiura[4, 5], Akira Uchiyama[4, 6], Ashif A. Fathnan[1], Hiroki Wakatsuchi[1, 4]\**

Dedication

M. Tashiro, K. Ide, K. Asano, Prof. S. Ishii, Prof. Y. Sugiura, Prof. A. Uchiyama, Dr. A. A. Fathnan, Prof. H. Wakatsuchi

*Corresponding author (wakatsuchi.hiroki@nitech.ac.jp)

Affiliations

[1] Department of Engineering, Nagoya Institute of Technology, Nagoya, Aichi, 466-8555, Japan

[2] Research Center for Materials Nanoarchitechonics (MANA), National Institute for Materials Science (NIMS), Tsukuba, Ibaraki, 305-0044, Japan

[3] Faculty of Pure and Applied Physics, University of Tsukuba, Tsukuba, Ibaraki, 305-8577, Japan

[4] Precursory Research for Embryonic Science and Technology (PRESTO), Japan Science and Technology Agency (JST), Kawaguchi, Saitama, 332-0012, Japan

[5] Graduate School of Science and Technology, Keio University, Yokohama, Kanagawa, 223-8522, Japan

[6] Graduate School of Information Science and Technology, Osaka University, Suita, Osaka, 565-0871, Japan

## Abstract

IoT sensors are crucial for visualizing multidimensional and multimodal information and enabling future IT applications/services such as cyber-physical space, digital twins, autonomous driving, smart cities, and virtual/augmented reality (VR or AR). However, IoT sensors need to be battery-free to realistically manage and maintain the growing number of available sensing devices. Here, we provide a novel sensor design approach that employs metasurfaces to enable multifunctional sensing without requiring an external power source. Importantly, unlike existing metasurface-based sensors, our metasurfaces can sense multiple physical parameters even at a fixed frequency by breaking classic harmonic oscillations in the time domain, making the proposed sensors viable for usage with limited frequency resources. Moreover, we provide a method for predicting physical parameters using the machine learning-based approach of random forest regression. The sensing performance was confirmed by estimating temperature and light intensity, and excellent determination coefficients larger than 0.96 were achieved. Our study affords new opportunities for sensing multiple physical properties without relying on an external power source or needing multiple frequencies, which markedly simplifies and facilitates the design of next-generation wireless communication systems.

## 1. Introduction

The ability to perceive and interpret information from several dimensions and modes is essential for advancing future information technologies.[1–3] Today, various sensors are used to detect diverse physical parameters, including temperature, light intensity, humidity, pressure, sound, angle, posture, pollution, and radiation, as shown in Figure 1a.[4] These sensors serve numerous purposes, such as optimizing power consumption, enhancing healthcare, preserving the environment, supporting agriculture, and ensuring security. Recently, they have been incorporated into wireless networks as Internet of Things (IoT) systems, which enable future cyber-physical space, digital twins, autonomous driving, smart cities, and virtual/augmented reality (VR or AR).[1,2,5] This trend is reflected in the global prevalence of IoT devices, specifically the substantial rapid annual growth rate of two billion or more devices per year. Nevertheless, the growing demand for IoT sensors has raised significant concerns about the management of a vast quantity of devices with limited human resources. More precisely, while these devices rely on batteries to establish communication with remote systems, it is not feasible to manually provide a new battery for every individual device. Therefore, battery-free or maintenance-free sensors are ideal for next-generation IoT systems.

Metasurfaces can serve as a viable option in this scenario for detecting physical quantities without the need for battery replacement.[6–10] Metasurfaces are artificially engineered structures that exhibit distinct behavior based on the properties of their subwavelength unit cells and the spectra of the incoming wave.[11–13] Metasurfaces exhibit a robust response to an incoming wave at a designed resonant frequency and can efficiently sense physical properties such as light intensity and temperature by incorporating vanadium dioxide, MEMS, thermistors, and/or photocells (i.e., photoresistors) into the metasurface unit cells (Figure 1b).[14–18] In such an approach, every identified physical property is associated with a single resonant frequency. Thus, two physical properties can be identified using two independent resonant frequencies, which implies that two physical parameters cannot independently be detected at a single frequency (Figure 1c). Moreover, in practice, the allocation of frequency resources is rigorously regulated.[19–22] Thus, the optimal scenario would include overcoming the frequency-domain restriction imposed by classic resonant mechanisms and detecting multiple physical quantities at only a single frequency. For this reason, this study proposes metasurface-based sensors that change scattering profiles depending on the physical properties of the surrounding environment even at the same frequency (Figure 1d and Figure 1e). Our sensors are specifically designed to detect light intensity and temperature via the integration of photocells and temperature-dependent capacitors. Importantly, however, these physical quantities are identified by using only a single frequency thanks to time-domain scattering changes with a machine learning methodology.[23,24] By altering the integrated circuit layout, the design concept of our metasurface-based sensors may be extended to detect additional physical properties. Thus, this study helps achieve maintenance-free and sustainable next-generation wireless communication systems.

## 2. Results and Discussion

### 2.1. Fundamental Design Theory

A key solution to detecting multiple physical quantities at the same frequency is in breaking the harmonized time-domain response. Toward this goal, time-varying metasurfaces have been intensively studied and exploited for wavefront control, which aids in designing reconfigurable intelligent surfaces (RISs),[25–27] nonreciprocal wave propagation,[28,29] and radio-frequency (RF) and optical devices.[30,31] Although most time-varying metasurfaces require external power sources as active metasurfaces,[26,32–34] passive and time-varying metasurfaces were recently proposed to change the electromagnetic response even at the same frequency in accordance with the duration of the incoming pulse;[31,35–37] such an approach is exploited in this study. In fact, these pulse-width-dependent metasurfaces, or so-called waveform-selective metasurfaces, rely on the well-known transients of classic direct-current (DC) circuits. More specifically, waveform-selective metasurfaces consist of a periodic conducting pattern and resonate at resonant frequencies, as seen in ordinary metasurfaces.[38,39] However, by introducing a set of four diodes as a diode bridge into the gap between conductor edges, the waveform of the incoming wave (the sine function in this study) is fully rectified (as in the waveform based on |sin|), generating an infinite set of frequency components in which most of the energy is concentrated around the zero frequency band. Therefore, transient phenomena can occur if reactive circuit components are included inside the diode bridge. Specifically, this study uses a capacitor connected to a resistor in parallel inside the diode bridge. Under this circumstance, the reflection from the metasurface is reduced during an initial period because the incoming energy is stored in the capacitor and dissipated with the parallel resistor. However, by increasing the incident pulse width, the capacitor is fully charged so that the incident wave is poorly absorbed and strongly reflected even at the same frequency.

In particular, this study exploits that the transient responses are characterized by time constants and steady-state resistance, which are associated with sensing circuit components that vary their circuit values in accordance with physical quantities. For instance, some capacitors are well known to change their capacitance due to temperature changes. Additionally, photocells show variable resistance values in accordance with the surrounding light intensity. Therefore, by incorporating these circuit elements as parallel capacitors and resistors inside diode bridges, transients (or time-varying responses) change depending on temperature and light intensity, which can be detected from scattering waves. Specifically, as explained in the literature[40] and Supporting Information (Supporting Note 1), the time constant of our metasurface $\tau$ is determined by

$$\tau = \frac{CR_C R_d}{R_C + R_d},\qquad(1)$$

where $C$ and $R_C$ represent the capacitance and resistance of the discrete components inside the diode bridge (i.e., the parallel capacitor and resistor). Additionally, $R_d$ denotes the resistance of the two diodes at the turn-on voltage. In particular, if $R_C \gg R_d$, $\tau$ is simplified to

$$\tau \sim CR_d,\qquad(2)$$

which indicates that the transition time is mostly changed by $C$, since $R_C$ is not adjustable. Additionally, in the steady state, the capacitor approaches an open circuit so that the reflecting state is related to the values of $R_C$ and $R_d$. Thus, $R_C$ can be exploited to control the steady-state response. Moreover, because $C$ and $R_C$ vary due to the temperature dependence of capacitors and the light-intensity dependence of photocells, our metasurface can detect temperature and light intensity in accordance with the reflected waveform. Note that while this study limits the multifunctional sensing capability to only two physical quantities, the proposed concept can be further extended to detecting additional physical quantities by introducing extra circuit components. For instance, the above capacitor-based circuit configuration can be integrated with an inductor-based circuit configuration to produce a reflectance peak, dip, or more advanced waveform, which can be associated with more than two circuit parameters in the time domain.[41,42]

## 2.2. Numerical Demonstration

Before experimental validation, we numerically show how metasurface-based sensors vary their time-domain response in accordance with circuit parameters. As seen in Figure 2a, the unit cells of our metasurface consist of a ground plane, a dielectric substrate (Rogers 3003), and conducting patches with minor trimming to deploy small conducting pads and form a diode bridge including a parallel $R_C C$ circuit ($R_C$ = 10 kΩ and $C$ = 1 nF). For the sake of simplicity, these simulations use ordinary capacitors and resistors for $C$ and $R_C$ instead of a temperature-dependent capacitor and a photocell, respectively. Detailed information for these simulations is provided in "Simulations" of the Experimental Section and the Supporting Information (Supporting Note 2), including the design parameters for the conducting geometry, substrate, and lumped circuit elements. Under these circumstances, the reflecting profiles of the metasurface for 50-ns short pulses and continuous waves (CWs) varied, as shown in Figure 2b. According to these simulation results, the metasurface significantly reduced the reflectance magnitude only for short pulses near 3.9 GHz, which is consistent with the pulse-width-dependent absorbing mechanism explained earlier. To further clarify this reflectance trend, $C$ and $R_C$ were varied, as shown in Figure 2c, which indicates that the transition time was shifted by increasing $C$ from 10 nF to 100 nF. Additionally, the steady-state reflectance decreased upon decreasing $R_C$ from 10 kΩ to 1 kΩ. Importantly, these two conclusions also support the aforementioned theoretical design and ensure that two independent physical quantities can be detected if they are associated with changes in $C$ and $R_C$. It is also noted that related results and information are provided in the Supporting Information (Supporting Note 2).

## 2.3. Experimental Demonstration

Based on the simulation results shown above, we fabricated and experimentally tested the metasurface-based sensor, as shown in Figure 3a. In this measurement, our metasurface was composed of 15 × 15 unit cells and placed on a programmable hot plate that arbitrarily and directly controlled the temperature of the metasurface sample. Additionally, a light source was positioned in front of the metasurface and controlled by pulse width modulation (PWM) signals. In addition to these thermal and light sources, incident signals were radiated by a standard horn antenna, and another horn antenna was used to receive the reflected waveform. Both the incident and reflected angles were set to 30 degrees, and the incident wave was a transverse electric (TE) polarized wave. Detailed information for the measurements is provided in "Measurement Samples" and "Measurement Method" of the Experimental Section and the Supporting Information (Supporting Note 2). Under these circumstances, the temperature of the metasurface and the surrounding light intensity were set to 22.1 degrees Celsius and 328 lux, respectively, which provided nearly identical values of $C$ and $R_C$ as those shown in Figure 2b (specifically, 10 nF and 10 kΩ). As a result, the frequency-domain profiles shown in Figure 3b demonstrated a relatively low transmittance for short pulses near 3.76 GHz despite the use of the same frequencies, which was consistent with the simulation results in Figure 2b. Note that the transmittance in Figure 3b was entirely lower than that in Figure 2b because the measurement was performed in open space to consider a realistic sensing environment, while the simulation was conducted with periodic boundaries as a simplified situation. Additionally, a minor frequency shift appeared in the measurement because of differences between the simulation and measurements, e.g., the incident angle and parasitic circuit parameters that only appeared in the measurement sample. However, despite these differences, the reflectance profile evidently varied in accordance with the incident waveform even during the measurements.

Next, we clarified how the time-domain response varied with changes in temperature and light intensity. In these measurements, we simplified the loaded circuits inside the diode bridges and used pairs of either temperature-dependent capacitors and fixed resistors or fixed capacitors and photocells, which facilitated the analysis of the temperature and light intensity dependences. First, pairs of temperature-dependent capacitors and fixed resistors (10 kΩ) were used with variable temperature values at 3.82 GHz, as shown in Figure 3c. According to these measurements, the transition time decreased with increasing metasurface temperature from 23.5 °C to 65.0 °C because the capacitance of the temperature-dependent capacitors decreased, which decreased the time constant. This result was consistent with the numerical simulation in Figure 2c and Equation (2). Additionally, when pairs of fixed capacitors (1 nF) and photocells were alternatively used within the diode bridges, the metasurface varied the steady-state reflectance, as shown in Figure 3d. Specifically, the reflectance decreased from -30.9 dB to -37.5 dB by increasing the light intensity from 3 LUX to 1970 LUX, which resulted in lowering the effective resistive component of the photocells and strongly absorbing the incident wave in the steady state.

Moreover, using both the temperature-dependent capacitors and the photocells, the metasurface was experimentally evaluated in Figure 3e. This measurement result also ensures that the changes in temperature and light intensity affect the reflectance profiles. For instance, by increasing the temperature from 40 to 70 degrees Celsius, the time constant of the metasurface was reduced so that the reflectance curves were shifted to a smaller time scale (i.e., comparing the solid curves with the dashed curves). Additionally, by increasing the light intensity from approximately 300 to 2000 LUX, the steady-state reflectance decreased from approximately -23 dB to -33 dB (see the difference between the black curves and the red curves). Therefore, by associating $C$ and $R_C$ with physical quantities, our metasurface design independently controlled the time constant and the steady-state response. This metasurface-based approach is also demonstrated using simpler versions of structures such as microstrips and one-dimensional metasurface lines in the Supporting Information (Supporting Note 3), which further validates the time-varying scattering effect even at the same frequency in accordance with temperature and light intensity.

## 2.4. Estimation of Physical Quantities

We further present an approach to estimate temperature and light intensity based on the physical quantities associated with $C$ and $R_C$. Although other approaches are potentially applicable for predicting these two physical quantities (e.g., use of theoretical equivalent circuit models[40,43–45]), we adopted a machine learning approach based on random forest regression.[24] Our estimation approach was composed of four steps. First, the measured time-domain reflectance profiles were divided into 40 segments of time on a log scale. Second, in each segment, an average reflectance value was obtained. Third, these average reflectance values were used as explanatory variables for the training data. Finally, based on the training data, the temperature and light intensity were estimated and compared to their actual values. We changed the number of training datasets and that of test datasets, while the total number of these datasets was fixed at 2290 (i.e., the ratio between the training datasets and the test datasets was changed within the entire dataset of 2290). Further details are provided in "Estimation Method" of the Experimental Section.

The corresponding estimation results are shown in Figure 4. When only 11 datasets were used as training datasets in Figure 4a and Figure 4b, the correlation between the estimated values and the original values was poor, with determination coefficients of 0.6456 and 0.5841 for temperature and light intensity, respectively. However, by increasing the number of training datasets to 458, these determination coefficients improved to 0.9861 and 0.9610 for temperature and light intensity, respectively. These results indicate that a proper number of datasets needs to be used for the training process, which is consistent with other reports on AI-based metasurface studies.[46–48] More importantly, these results validate that our metasurface-based sensors can be used to estimate physical quantities at the same frequency. The Supporting Information (Supporting Note 4) provides additional results, including the use of a different estimation method instead of random forest regression.

## 2.5. Discussion

Our metasurface-based sensing approach was experimentally validated to be capable of estimating more than one physical quantity by breaking the harmonic oscillation of metasurfaces in the time domain and using only one frequency component. Our approach is rational because the use of frequency resources is strictly determined in practice to avoid electromagnetic interference issues.[19–21] This approach may be applicable to one of the industrial, scientific, and medical (ISM) bands, where frequency resources are readily available without rigorous license issues for using radio-frequency (RF) waves. Moreover, our approach is useful for managing and designing future IoT systems. Conventionally, IoT sensors were designed to obtain multidimensional and multimodal information for realizing next-generation IoT systems, including cyber-physical space, digital twins, autonomous driving, smart cities, and VR/AR.[1,2,5] However, this conventional approach requires regular replacement of internal batteries to maintain communication with external internet/cloud systems. Because the number of IoT sensors is increasing rapidly, maintaining all IoT sensors manually and replacing their batteries will no longer be realistic soon. Here, our metasurface-based approach does not require the use of batteries but permits the sensing of multiple physical quantities, addressing an emerging issue in the design of future IoT systems. In particular, while this study was limited to sensing two physical quantities as a proof of concept, our approach can be extended to sensing more than two physical quantities by adding additional circuit components that react with other physical quantities and characterize the time-domain response of metasurfaces.[37,41,42]

For more practical use as IoT sensors, further improvements are required for our metasurface-based sensors. For instance, the metasurfaces demonstrated in this study rely on commercial diodes that require a large amount of input power to rectify incoming signals and vary the time-domain response associated with physical quantities. Therefore, reducing the power level by using customized low-power diodes helps design battery-free metasurface-based multifunctional sensors in an energy-efficient manner. Additionally, the successful implementation of our approach in realistic environments depends on how the relationships between reflected (or scattered) waveforms are associated with physical quantities. This issue was addressed in this study by exploiting random forest regression, and other machine learning techniques can potentially be exploited to efficiently predict physical quantities, as shown in the Supporting Information (Supporting Note 4). Finally, connecting our sensing approach to IoT/cloud systems is another important issue that is not fully demonstrated in this study. In particular, real-time feedback is needed for IoT systems in cyber-physical space, smart cities, autonomous driving, farming, and healthcare,[49–52] where wireless networks are expected to reduce the estimation time for physical quantities by using, for instance, simplified learning models and fast-speed calculation approaches.

## 3. Conclusion

In conclusion, we present a metasurface-based sensor design that achieves multifunctional sensing without needing multiple frequencies or an external power supply. Our metasurface showed variable reflectance profiles in the time domain, which were independently determined by lumped circuit parameters that were responsive to the two physical quantities of interest, specifically, temperature and light intensity. Additionally, we introduced an approach to estimate the temperature and light intensity from the reflected waveform of the metasurface by using random forest regression. As a result, the temperature and light intensity were successfully detected with determination coefficients of 0.9861 and 0.9610, respectively. Our study affords new possibilities for sensing multiple physical quantities without needing an external power supply or several frequencies, which facilitates the design of next-generation wireless communication systems.

## 4. Experimental Section

*Simulations*: Numerical simulations were performed using a co-simulation method in ANSYS Electronics Desktop (version R2) 2022. In this method, metasurfaces were modeled in an electromagnetic solver (HFSS). Importantly, all of the discrete circuit components were replaced with lumped ports. The scattering parameters of the metasurfaces were then used in a circuit simulator (Circuit) as circuit models. In these circuit simulations, lumped ports were connected to the actual circuit components used, which was equivalent to directly including the circuit components in electromagnetic simulations. However, this co-simulation method facilitated the entire simulation process for readily obtaining the final simulation results compared to stand-alone electromagnetic simulation approaches.[53] Short-pulse simulations (as shown by the red curve in Figure 2b) were conducted using 50-ns pulses. In this case, the entire reflected energy was compared with the entire incident energy to calculate the reflectance. Additionally, CW simulations (as seen in the black curve of Figure 2b) were performed using the harmonic balance approach, where the steady-state response was directly obtained. Here, we calculated the reflectance by dividing the reflected energy by the incident energy during 2 cycles. Moreover, we also calculated the transient reflectance in the time domain (as shown in Figure 2c). In this case, the time-varying reflectance was obtained by calculating the moving average of the reflected energy for 250 ns with discretized 100-ps time steps and comparing it with the moving average of the incident energy. The detailed design parameters of the simulation models are given in the Supporting Information (Supporting Note 2).

*Measurement Samples*: Our metasurface measurement samples were composed of a ground plane, a dielectric substrate (Rogers 3003), and a periodic array of square conducting patches with minor trimming to connect discrete circuit components. These design parameters are fully summarized in the Supporting Information (Supporting Note 2). The diodes used were provided by Avago (HSMS-286x series). The temperature-dependent capacitors and photocells were produced by Murata Manufacturing Co. (RDEF51H013Z0P1H03B) and Luna Innovations (NSL-19M51), respectively.

*Measurement Method*: Although detailed measurement setups are illustrated in the Supporting Information (Supporting Note 2), to characterize frequency-domain profiles (as shown in Figure 3b), we used not only a vector network analyzer (VNA) (Keysight Technologies, N5249A) but also an amplifier (Ophir, 5193RF) to sufficiently increase the input power level and turn on the diodes loaded on the metasurfaces. For time-domain profiles (e.g., those shown in Figure 3c to Figure 3e), we used a signal generator (Anritsu, MG3692C) as a signal source. Similarly, the abovementioned amplifier was used to ensure that the input power level was sufficiently large. Additionally, an isolator was used to protect the amplifier and the signal generator from excessive reflection. Part of the incident wave was sent to an oscilloscope (Keysight, DSOX6002A), while most of the energy was radiated to the metasurfaces through a standard horn antenna (Schwarzbeck Mess-Elektronik, BBHA9120D). Importantly, the surrounding light intensity and the temperature of the metasurfaces were controlled by a light source (Safego, C36W-FL) and a programmable hot plate (AS ONE, ND-2A). These light and heat sources were arbitrarily controlled by pulse width modulation (PWM) and proportional-integral-derivative (PID) control, respectively. Due to these two sources, the metasurfaces varied the reflected waveforms that were received by another horn antenna and measured by the oscilloscope. As mentioned in the above "Simulations" subsection, the transient reflectance varying in the time domain was obtained by comparing the reflected energy with the incident energy.

*Estimation Methods*: The reflected waveform was used to estimate the actual temperature and light intensity. This study used Python program codes based on random forest regression to obtain the results in Figure 4. Here, we used a built-in function of Phyton (specifically, "split") to pick up training datasets at random from all of the measurement datasets (2290 in total), while the remaining datasets were used as test datasets. For instance, in Figure 4a and Figure 4b, 11 datasets were selected at random as training datasets, while the remaining 2279 datasets (=2290-11) were used as test datasets. In Figure 4c and Figure 4d, the number of training datasets was increased to 458, while that of the test datasets was reduced to 1832. In the Supporting Information (Supporting Note 4), the relationship between the number of training datasets and that of test datasets was varied to demonstrate how these numbers influence the estimation performance. Ridge regression[54] was also alternatively applied to estimate the temperature and light intensity in the Supporting Information (Supporting Note 4).

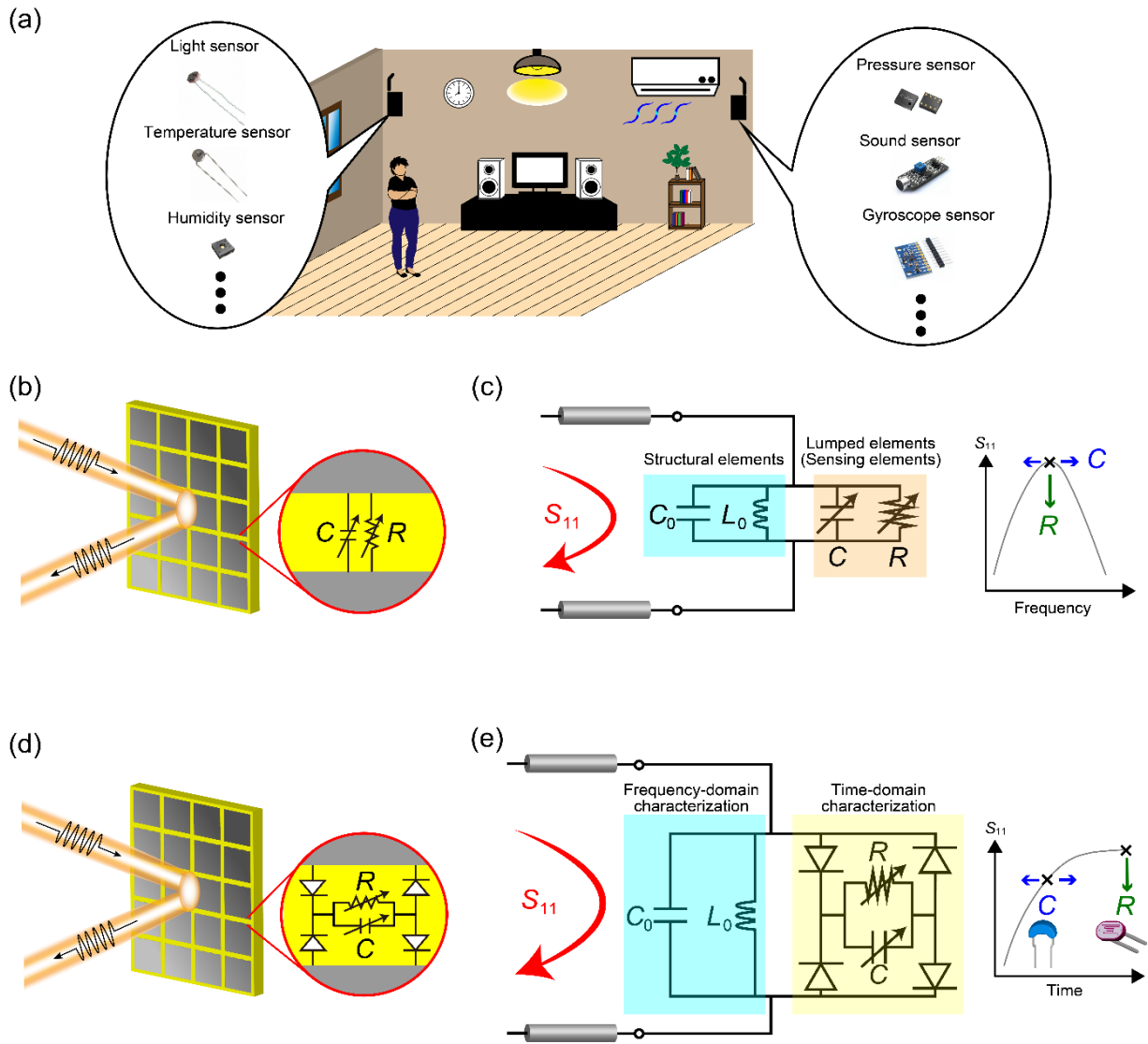

**Figure 1.** Concept underlying the metasurface-based sensors. (a) A variety of sensors that are used in daily life. (b) Conventional metasurface-based sensor design using frequency-domain scattering profiles that address the issue of regular battery replacement and (c) a corresponding equivalent circuit model with its scattering characteristics. Two circuit values (two physical properties) cannot independently be detected at a single frequency. (d) Proposed metasurface-based sensor design using time-domain scattering profiles that additionally address the restriction of limited frequency resources and (e) a corresponding equivalent circuit model with its scattering characteristics. Even at the same frequency, two physical properties can independently be detected in the time domain by using diode bridges with sensing circuit elements such as photocells and temperature-dependent capacitors. This sensor design breaks harmonized oscillations and attains a large degree of freedom to detect multiple physical parameters.

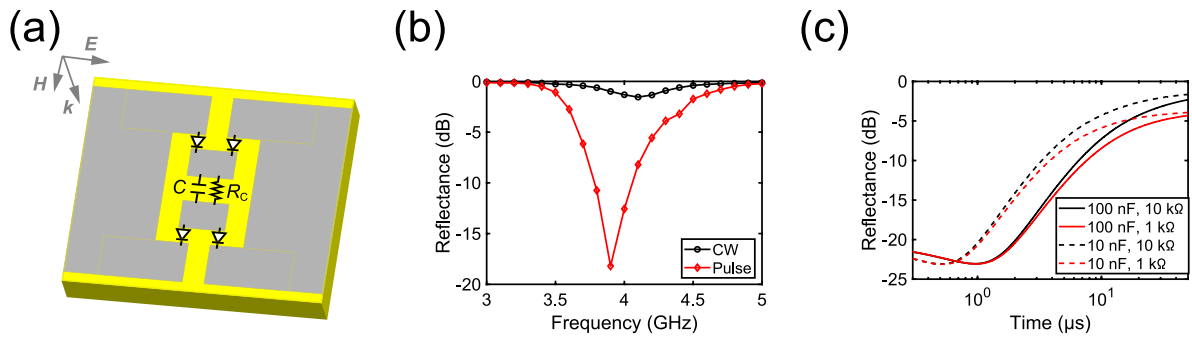

**Figure 2.** Numerically simulated reflectance profiles. (a) Realistic unit cell model with periodic boundaries applied for the incident $E$ and $H$ directions. (b) Frequency-domain reflectance. $C$, $R_C$, and the input power were set to 1 nF, 10 kΩ, and 0 dBm, respectively. (c) Time-domain reflectance at 3.9 GHz with various $C$ and $R_C$ values.

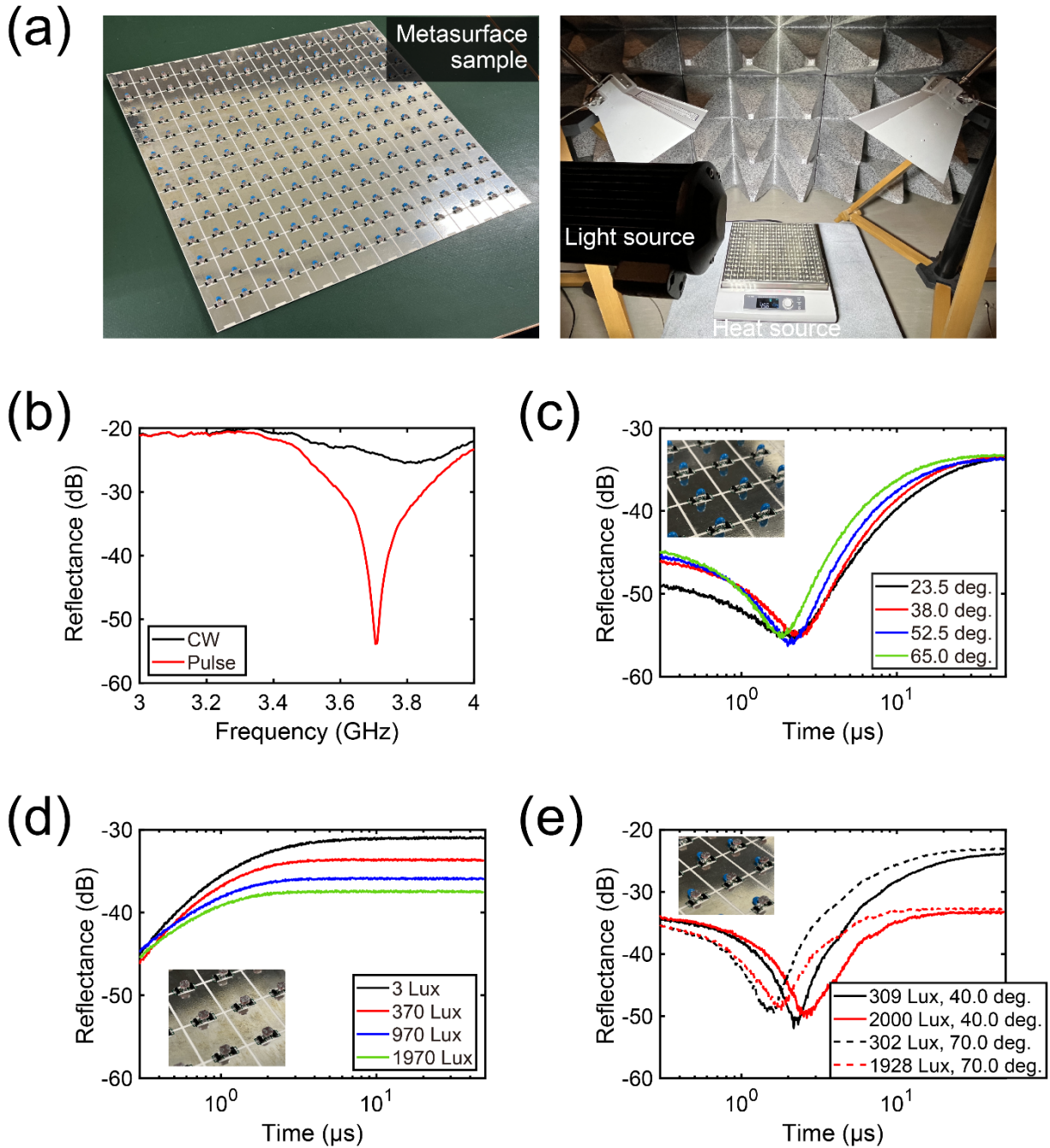

**Figure 3.** Experimental validation. (a) Measurement sample and measurement system using light and heat sources in free space. (b) Frequency-domain reflectance. The temperature, light intensity, and input power were set to 22.1 degrees Celsius, 328 LUX, and 30 dBm, respectively. (c-e) Time-domain reflectance of the metasurface-based sensor using (c) temperature-dependent capacitors and fixed resistors (10 kΩ), (d) fixed capacitors (1 nF) and photocells, and (e) both temperature-dependent capacitors and photocells. The frequency was set to optimal values to increase the time-domain variation in (c-e) (specifically, 3.82 GHz in (c) and (d) and 3.76 GHz in (e)).

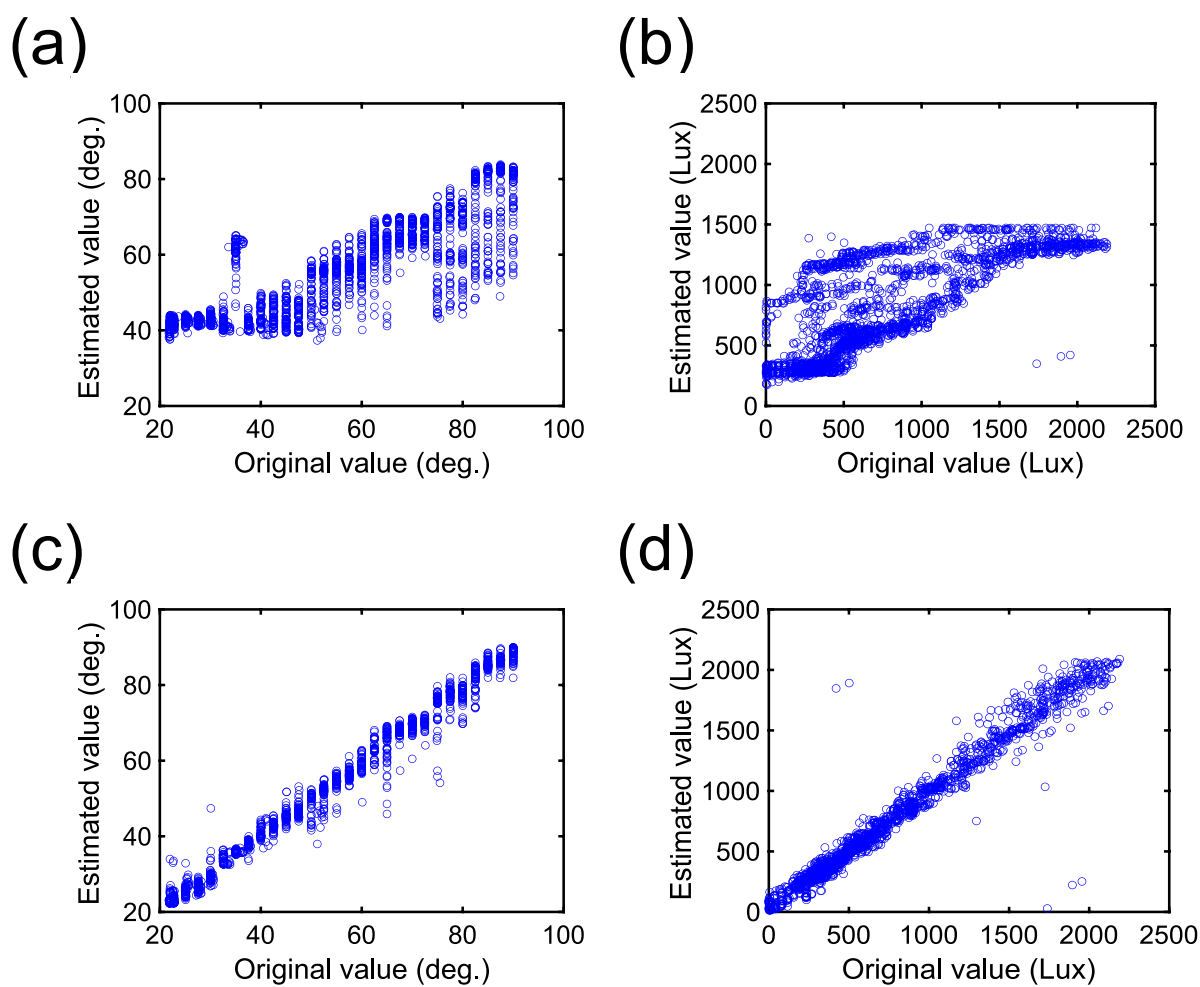

**Figure 4.** Simultaneous estimation of temperature and light intensity from reflected waveforms. (a, b) Use of 11 training datasets for (a) temperature and (b) light intensity estimation. (c, d) Use of 458 training datasets for (c) temperature and (d) light intensity estimation. The determination coefficients of (a) to (d) were 0.6456, 0.5841, 0.9861, and 0.9610, respectively. Additional results are provided in the Supporting Information (Supporting Note 4).

**Acknowledgements**

This work was supported in part by the Japan Science and Technology Agency (JST) under Fusion Oriented Research for Disruptive Science and Technology (FOREST) and under Precursory Research for Embryonic Science and Technology (PRESTO) Nos. JPMJPR193A, JPMJPR1932, and JPMJPR2134.

Table of Contents

Masaya Tashiro, Kosuke Ide, Kosei Asano, Satoshi Ishii, Yuta Sugiura, Akira Uchiyama, Ashif A. Fathnan, Hiroki Wakatsuchi

**Metasurface-Enabled Multifunctional Single-Frequency Sensors without External Power**

# Supporting Information

**Metasurface-Enabled Multifunctional Single-Frequency Sensors without External Power**

*Masaya Tashiro, Kosuke Ide, Kosei Asano, Satoshi Ishii, Yuta Sugiura, Akira Uchiyama, Ashif A. Fathnan, Hiroki Wakatsuchi\**

**Supporting Note 1: Simplified Equivalent Circuit Models for Metasurfaces**
**Supporting Note 2: Design of Simulation Models and Measurement Samples**
**Supporting Note 3: Simplified Test Models and Prototypes**
**Supporting Note 4: Estimation of Physical Quantities**

**Supporting Note 1: Simplified Equivalent Circuit Models for Metasurfaces**

This section explains how the response of our metasurfaces can be represented by simplified equivalent circuit models. As shown in Figure 2, our metasurfaces had periodically metalized patches that were connected by diode bridges, including RC circuits. In this case, the frequency-domain response can be represented by equivalent circuit components.[44,45] Additionally, the time-domain response can also be represented by equivalent circuit models under two assumptions.[40] First, the incoming waveform following a sine function is fully rectified by diode bridges to produce a different waveform based on a modulus of sin, generating an infinite number of frequency components. However, most of the converted energy is at zero frequency according to the Fourier expansion of |sin|.[37] Therefore, the circuit configuration between the periodically metalized patches was assumed to be biased by a DC voltage source instead of an alternating current (AC) source. Second, because the diode bridges cannot readily be approximated by simple circuit models due to their complicated nonlinear behavior, we assumed that the diode bridges were represented by fixed resistors that had the equivalent resistive component at the turn-on voltage extracted from the *I-V* characteristic of the diodes. Although this approach neglects the input power dependence of the effective resistive component of the diodes, as long as the incident power is sufficiently large, the time-domain characteristics can be closely approximated by the equivalent circuit model in Figure S1a,[40] as is leveraged in our study. Under these circumstances, currents and voltages are obtained by the following set of equations

$$\begin{cases} i_C(t) = \dfrac{dq(t)}{dt} \\ R_C i_R(t) = \dfrac{1}{C} \displaystyle\int i_C(t) dt \\ R_d\{i_R(t) + i_C(t)\} + R_C i_R(t) = E_0, \end{cases} \qquad \textbf{(S1)}$$

where $i_C$, $i_R$, $q$, and $t$ represent the currents at $C$ and $R_C$, the charges stored in $C$, and time, respectively. $E_0$ is the DC voltage applied across the entire circuit. By solving these equations, the capacitor voltage $v_C$ is obtained by

$$v_C(t) = \frac{R_C}{R_C + R_d} E_0 \left( 1 - e^{-\frac{t}{\tau}} \right), \qquad (S2)$$

where $\tau$ is the time constant shown in Equation (1) or Equation (2) (i.e., $\tau \sim CR_d$). Therefore, Equation (1) provides a good estimate of how quickly the time-domain response of our metasurface reaches the steady state. Additionally, according to Figure S1b, the current flowing into the entire circuit in the steady state is determined by $R_d$ and $R_C$ because $C$ is fully charged and becomes an open circuit in the steady state. For these reasons, the transition of the reflectance profiles of our metasurfaces can be controlled by $C$, while the steady-state response is tuned by $R_C$. By varying these two parameters with temperature-dependent capacitors and photocells, physical quantities (i.e., temperature and light intensity) were estimated in our study.

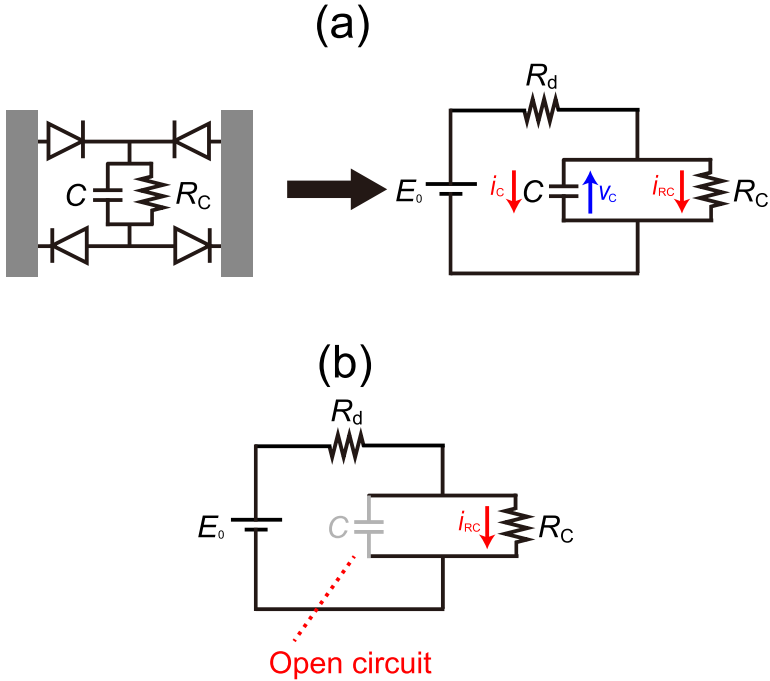

**Figure S1.** Equivalent circuit models for metasurfaces. (a) The circuit configuration between periodically metallized patches and simplified equivalent circuit for representing the time-domain characteristics of the metasurfaces. (b) Another equivalent circuit model representing the steady-state response.

**Supporting Note 2: Design of Simulation Models and Measurement Samples**

This section provides detailed information on the design parameters of the simulation models and measurement samples used. First, the simulation model demonstrated in Figure 2 used the design parameters and the circuit values shown in Figure S2, Table S1, Table S2, and Table S3. The same conditions were applied to our measurement samples. These samples were tested using the measurement systems shown in Figure S3 to obtain frequency-domain and time-domain profiles. Figure S3b shows the pulse-width-modulation-controlled LED and the hot place that were used in our measurements.

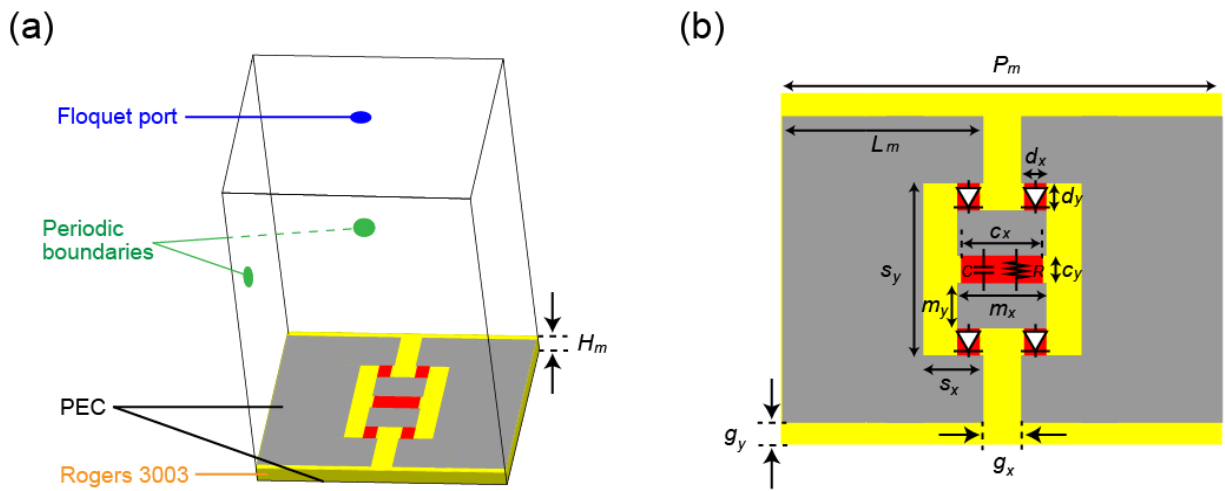

**Figure S2.** The simulation model used in Figure 2. (a) Entire model and (b) detailed dimensions. The design parameters and circuit values are presented in Table S1, Table S2, and Table S3.

**Table S1.** The design parameters used for the simulation model of Figure S2 (i.e., Figure 2).

| Parameter | Value (mm) |
|---|---|
| $P_m$ | 18 |
| $H_m$ | 1.52 |
| $L_m$ | 8.5 |
| $g_x$ | 1 |
| $g_y$ | 0.5 |
| $d_x$ | 0.5 |
| $d_y$ | 1.3 |
| $c_x$ | 1 |
| $c_y$ | 2 |
| $m_x$ | 1.7 |
| $m_y$ | 7.6 |
| $s_x$ | 1.7 |
| $s_y$ | 7.6 |

**Table S2.** The circuit values used for the simulation model of Figure S2 (i.e., Figure 2).

| Parameter | Value |
|---|---|
| Resistance:$R$ | **10 kΩ** |
| Capacitance:$C$ | 1 nF |
| Diode | HSMS-286x |

**Table S3.** SPICE parameters used for the simulation model of Figure S2 (i.e., Figure 2).

| Parameter | Value |
|---|---|
| $I_{BV}$ | 1e-15 A |
| $I_S$ | 5e-8 A |
| $N$ | 1.08 |
| $R_S$ | **6 Ω** |
| $V_B$ | 7 V |
| $C_J$ | 0.18 pF |
| $M$ | 0.5 |
| $P_B$ | 0.65 V |
| $E_G$ | 0.69 eV |

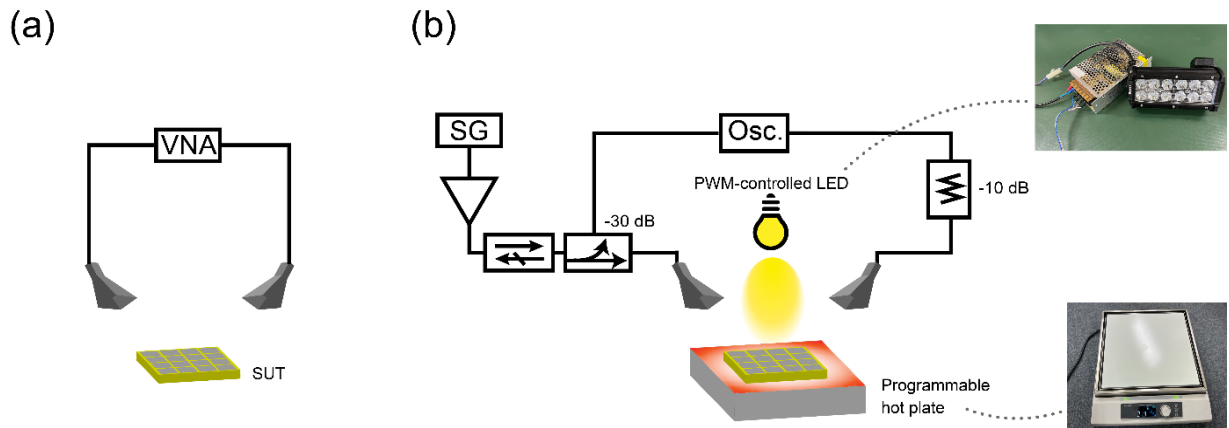

**Figure S3.** The measurement systems used. The systems for (a) frequency-domain profiles and (b) time-domain profiles.

**Supporting Note 3: Simplified Test Models and Prototypes**

Compared to the simulation models and measurement samples in Figure 2 and Figure 3, this section presents the results of simplified test models and prototypes to more clearly demonstrate the proposed concept of multifunctional sensing from scattering parameters. First, a simplified simulation model based on a microstrip is shown in Figure S4. As seen in ordinary microstrips,[55] our model was designed to transmit an incoming signal at a wide range of frequencies. However, a gap was made in the middle of the model and connected to an LC resonator ($C_{add}$ and $L_{add}$) to preferentially reject incoming signals at a design resonant frequency. More importantly, due to the presence of an additional diode bridge containing a parallel RC circuit ($C_S$ and $R_S$), incident waves were prevented depending not only on the frequency but also on the pulse duration, as seen in the waveform-selective mechanism demonstrated in Figure 2. The detailed design parameters and circuit values are shown in Table S4 and Table S5. The SPICE parameters were the same as those adopted in Table S3.

The simulated transmittance results are shown in Figure S5. According to these results, the transient transmittance varied even at the same frequency, as shown in Figure 2. In addition, the transition of the transmittance was clearly shifted by changing $C_S$, although $R_S$ also slightly contributed to varying this transition. Moreover, the steady-state transmittance was determined by $R_S$ because $C_S$ became an open circuit in the steady state and was independent of the steady-state transmittance.

(a)                                                                    (b)

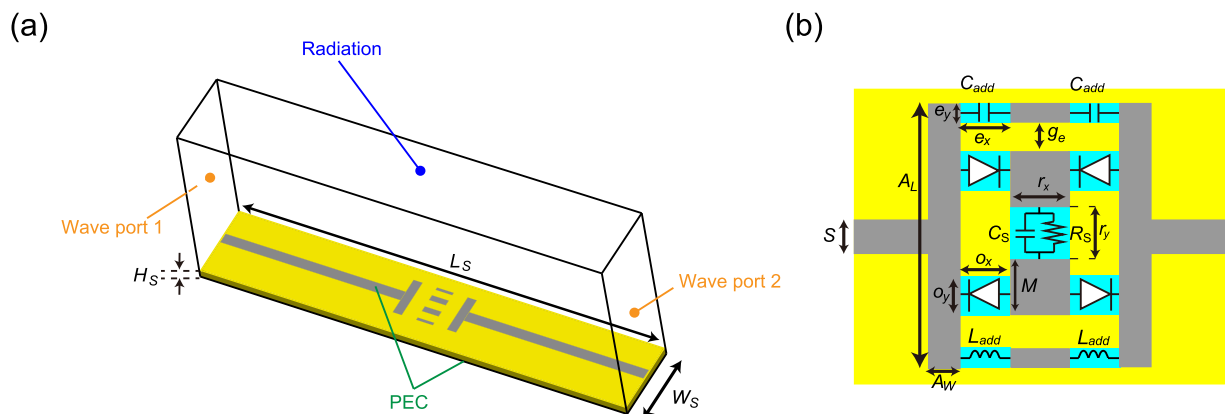

**Figure S4.** Microstrip model including the concept of waveform-selectivity. (a) The entire model and (b) a gap connected by an LC resonator and a waveform-selective circuit. $C_{add}$ and $L_{add}$ were fixed at 0.4 pF and 2.7 nH, respectively, while $C_S$ and $R_S$ were varied to clarify the waveform-selective response of the microstrip model. The design parameters and circuit values used are shown in Table S4 and Table S5.

**Table S4.** The design parameters used for the simulation model of Figure S4.

| Parameter | Value (mm) |
|---|---|
| $L_S$ | 131 |
| $W_S$ | 31 |
| $H_S$ | 1.52 |

**Table S5.** The circuit values used for the simulation model of Figure S4.

| Parameter | Value (mm) |
|---|---|
| $S$ | 3.05 |
| $A_L$ | 11 |
| $A_W$ | 1 |
| $e_x$ | 1 |
| $e_y$ | 1 |
| $o_x$ | 1 |
| $o_y$ | 1 |
| $r_x$ | 2 |
| $r_y$ | 1 |
| $M$ | 2 |
| $g_e$ | 2 |

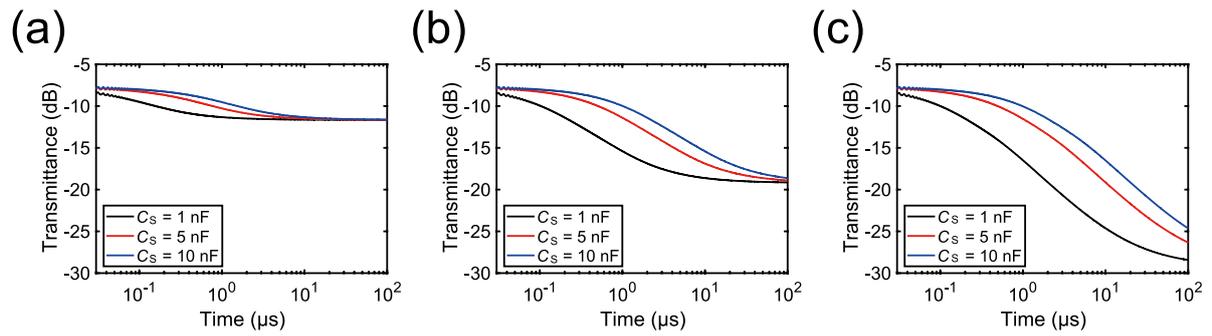

**Figure S5.** Simulated transient transmittance results of the microstrip model of Figure S4. The results using various capacitances $C_S$ and $R_S$ = (a) 0.1 kΩ, (b) 1 kΩ, and (c) 10 kΩ at 2.5 GHz with 10 dBm.

Based on the simulation results of Figure S5, we fabricated simplified measurement samples comprising a microstrip, an LC resonator, and a diode bridge including an RC circuit. As shown in Figure S6, we prepared three types of measurement samples. The first sample contained a temperature-dependent capacitor and a fixed resistor (10 kΩ) as the RC circuit. The second sample alternatively contained a fixed capacitor (1 nF) and a photocell. The third sample contained both a temperature-dependent capacitor and a photocell. All of the samples had SMA jack print-circuit-board (PCB) edge-mount connectors to send and receive signals via coaxial cables that were connected to a signal generator (Anritsu, MG3692C) and an oscilloscope (Keysight, DSOX6002A), as shown in Figure S7. To vary the temperature of the measurement sample, unlike the measurement in Figure 3, where many temperature-dependent capacitors were soldered to measurement samples, we used a hair dryer as a heat source in the measurement of Figure S7 because only one component had a temperature dependence that could be more readily and quickly controlled by the hair dryer. In contrast, the light intensity dependence was evaluated using the same method as that adopted in Figure 3.

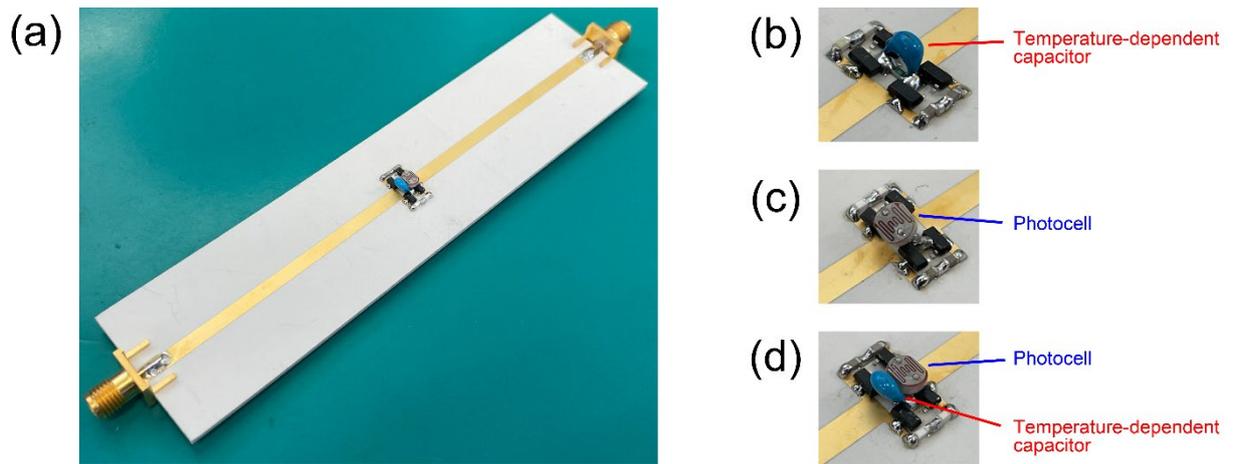

**Figure S6.** Microstrip-based measurement samples. (a) An entire image. The measurement samples using (b) a pair of a temperature-dependent capacitor and a fixed resistor (10 kΩ), (c) a pair of a fixed capacitor (1 nF) and a photocell, and (d) a pair of a temperature-dependent capacitor and the photocell.

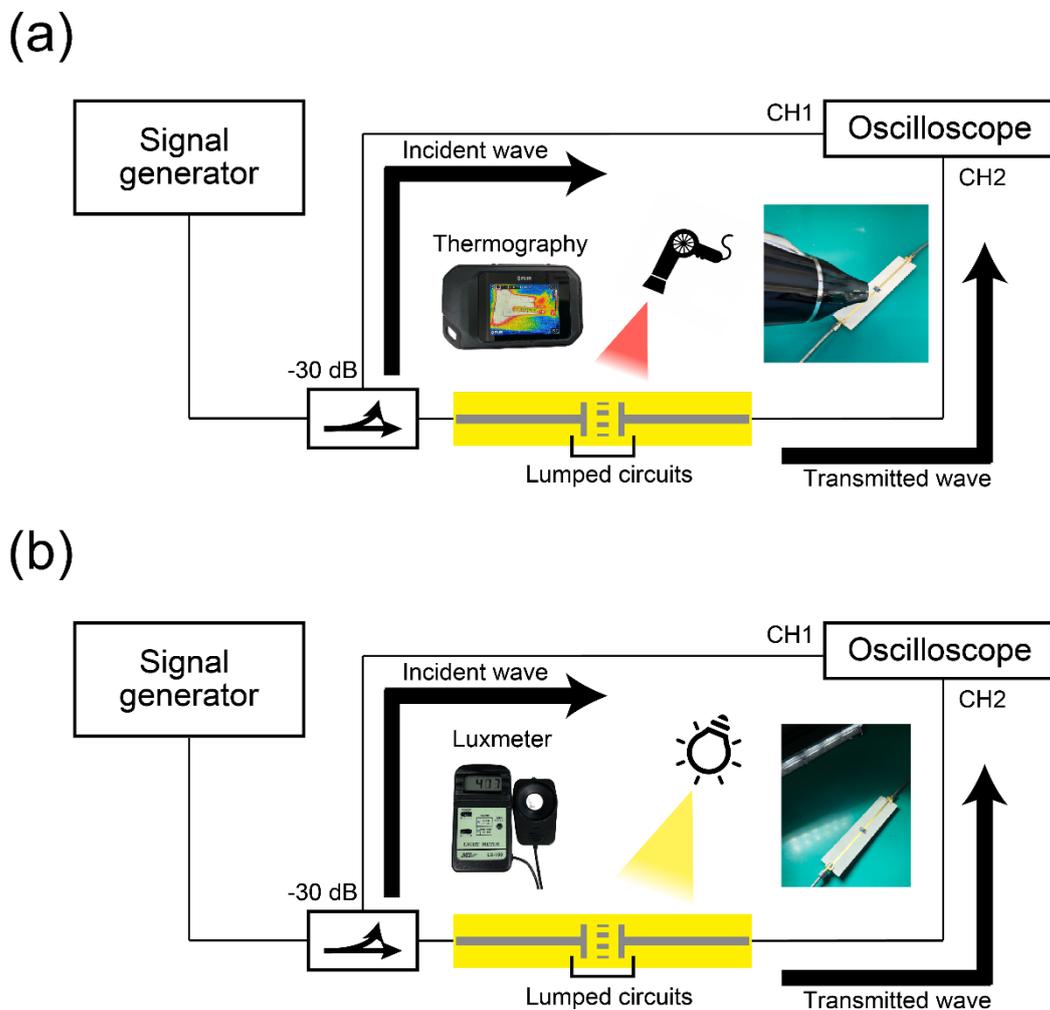

**Figure S7.** The measurement setups adopted for the microstrip-based samples shown in Figure S6. The setups for varying (a) the temperature and (b) the light intensity of the measurement samples.

The measurement results of the microstrip-based measurement samples are shown in Figure S8. First, the sample using a pair of a temperature-dependent capacitor (the same as those adopted in Figure 3) and a fixed resistor (10 kΩ) exhibited varying transmittance at the same frequency of 2.25 GHz with a 10-dBm input power. In particular, by increasing the temperature of the capacitor from 19.8 to 53.4 degrees Celsius, the transition of the varying transmittance shifted to a smaller time scale. This was because the temperature increase led to a decrease in the capacitance of this component as well as the time constant of the entire circuit configuration (refer to Equation (1) or Equation (2)). Next, the measurement sample was evaluated using a pair of a fixed capacitor (1 nF) and a photocell (the same component as the ones adopted in Figure 3), as shown in Figure S8b. As a result, the steady-state transmittance increased with increasing light intensity around the measurement sample. In this case, the resistive component of the photocell decreased, which permitted more incident energy to be transmitted without being dissipated within the circuit configuration in the steady state. These temperature-dependent capacitors and photocells were both included within the third measurement sample shown in Figure S8c. In this case, the measurement sample showed both a horizontal shift in the transient transmittance and a vertical shift in the steady-state transmittance in accordance with the changes in temperature and light intensity, respectively. Thus, these simplified measurement samples also support the change in the scattering parameters by varying the temperature and the light intensity.

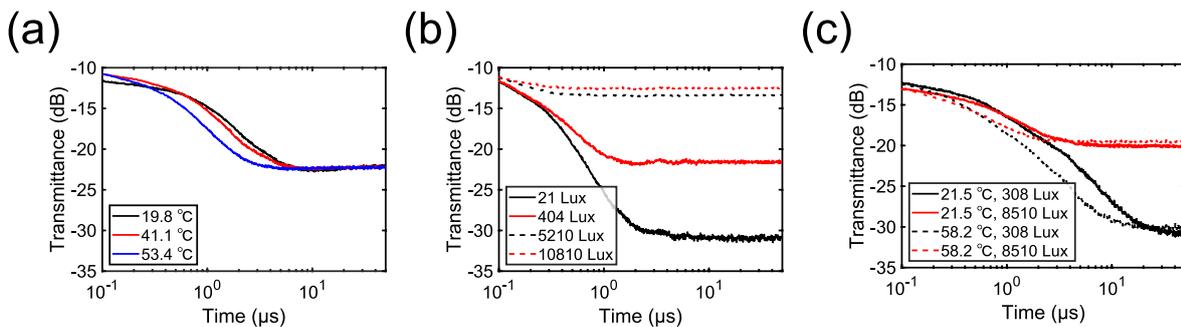

**Figure S8.** Measured transient transmittance results of the microstrip-based samples shown in Figure S6. The results using (a) a pair of a temperature-dependent capacitor and a fixed resistor (10 kΩ), (b) a pair of a fixed capacitor (1 nF) and a photocell, and (c) a pair of both the temperature-dependent capacitor and the photocell at 2.25 GHz with 10 dBm. The measurement samples of (a) to (c) correspond to Figure S6b, Figure S6c, and Figure S6d, respectively.

Next, we extended the concept of multifunctional sensing from the simplified models and samples in Figure S4 and Figure S6 to a one-dimensional line model summarized in Figure S9 and Table S6. In this model, we used a grounded monopole transmitter and receiver, as shown in Figure S10, which demonstrates that our monopoles were designed to effectively radiate and receive signals near 5 GHz. In particular, by deploying a metasurface line including diode bridges and RC circuits, a surface wave can be strongly guided depending on the waveform or the pulse width.[22] In this study, we additionally controlled the waveform-dependent impedance of the metasurface line by replacing the RC circuits with pairs of temperature-dependent capacitors and photocells (e.g., Figure S6). However, this one-dimensional line model simply adopts fixed capacitors and resistors with variable circuit constants to characterize the waveform-selective response of the simulation model at the same frequency, as shown earlier (e.g., Figure S5).

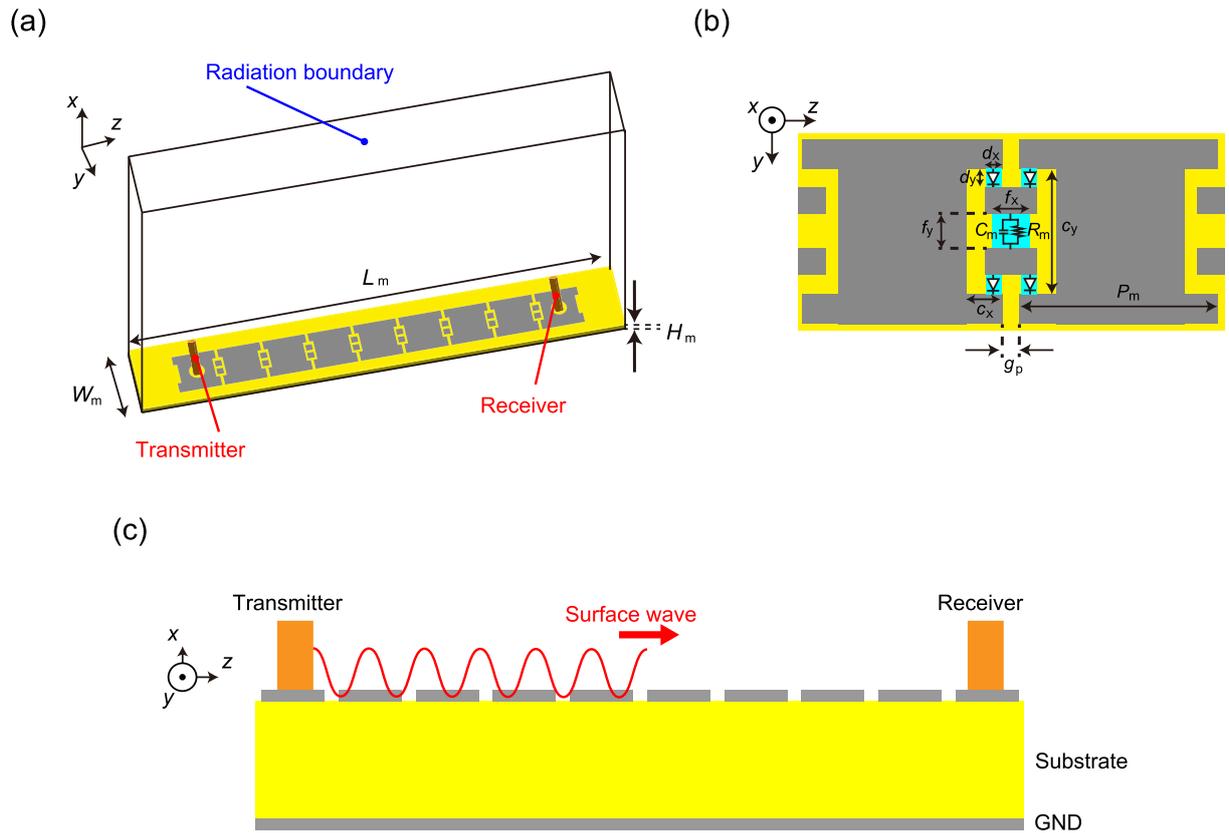

**Figure S9.** One-dimensional metasurface line model. (a) Angled view of the entire model comprised of 9 unit cells. (b) Detailed unit cell design. $C_m$ and $R_m$ were varied to clarify the waveform-selective response of the one-dimensional metasurface line model. (c) Cross-sectional view of the model. The design parameters used are shown in Table S6.

**Table S6.** The design parameters used for the one-dimensional metasurface line model of Figure S9.

| Parameter | Value (mm) |
|---|---|
| $H_m$ | 1.52 |
| $W_m$ | 322 |
| $L_m$ | 124 |
| $P_m$ | 17 |
| $g_p$ | 1 |
| $a_x$ | 1.7 |
| $a_y$ | 7.6 |
| $b_x$ | 0.5 |
| $b_y$ | 1.3 |
| $f_x$ | 1 |
| $f_y$ | 2 |

(a)     (b)

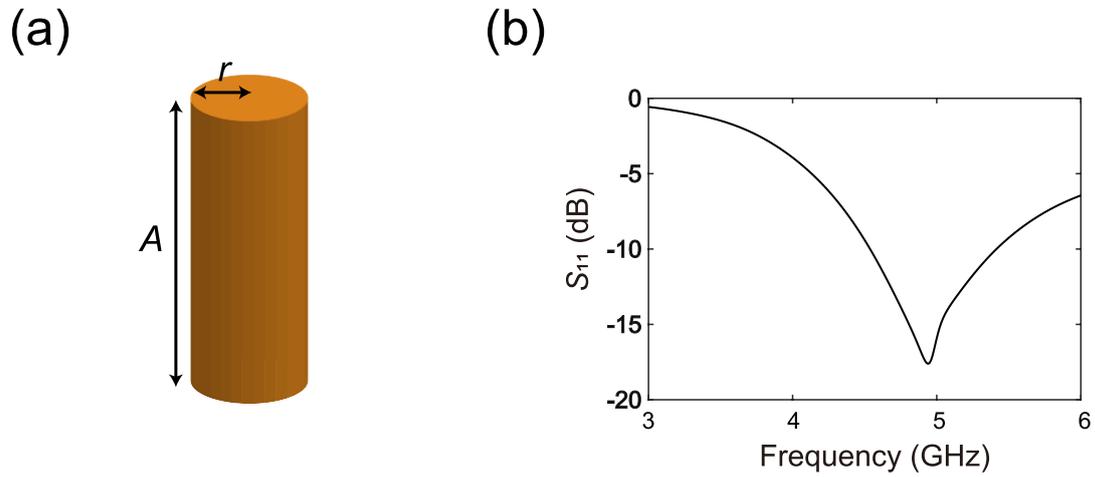

**Figure S10.** The grounded monopole antenna model used in Figure S9. (a) The model and (b) its $S_{11}$. $A$ and $r$ were set to 14 mm and 0.5 mm, respectively.

The simulation results of the one-dimensional metasurface line model (i.e., Figure S9) are plotted in Figure S11, which also shows variable transmittance depending on the circuit values of the embedded RC circuits. For instance, by increasing the resistance from 0.1 kΩ to 10 kΩ, the steady-state transmittance increased to a greater extent. In this case, the circuit part, including the diode bridges and the RC circuits, became an open circuit such that the metasurface was incapable of absorbing the energy of the incoming surface wave that was effectively transmitted to the receiver. Additionally, by increasing the capacitance of the RC circuits, the transient response of the metasurface was shifted to a larger time scale, as the time constant was increased by the capacitance (see Equations (1) and (2)[40]). Note that compared to the results of the microstrip model (i.e., Figure S5), which gradually decreased the transmittance, the simulated transmittance results of Figure S11 gradually increased. This occurred because the metasurface model in Figure S11 behaved as an absorber for short pulses while transmitting long pulses even at the same frequency.[22] However, the microstrip model did not exhibit a stop band as long as the capacitor included within the diode bridge was not fully charged. Therefore, these transmittance profiles were essentially different, although both models showed time-varying scattering profiles in accordance with the incident pulse width.

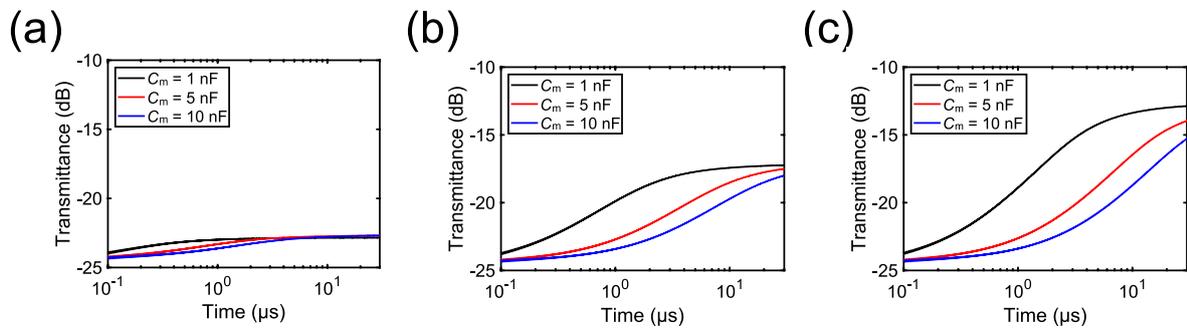

**Figure S11.** Simulated transient transmittance results of the one-dimensional metasurface line model of Figure S9. The results using various capacitances $C$s and $R$s = (a) 0.1 kΩ, (b) 1 kΩ, and (c) 10 kΩ at 4.9 GHz with 13 dBm.

Next, we fabricated measurement samples of the one-dimensional metasurface line model shown in Figure S9 and measured their transient transmittance. As shown in Figure S12, we used a signal generator and an oscilloscope to obtain transmittance profiles in the time domain. The measurement samples were connected to coaxial cables via grounded monopole antennas so that the transmittance between them varied depending on the impedance of the metasurface line. Here, we prepared three types of measurement samples: the first sample using temperature-dependent capacitors and fixed resistors (10 k$\Omega$), the second sample using fixed capacitors (1 nF) and photocells, and the third sample using both temperature-dependent capacitors and photocells. These samples were tested with 13-dBm signals at 4.81 GHz, where time-domain changes were most effectively observed, as shown in Figure S13. These results also support the same conclusion as those drawn in the simulation results of Figure S11 and the measurement results of the microstrip-based samples of Figure S8. For instance, Figure S13 shows that the time slot for exhibiting time-varying transmittance shifted to a smaller time scale by increasing the temperature of the temperature-dependent capacitors, as this temperature increase led to a decrease in the corresponding capacitance value and time constant (Figure S13a). In addition, by increasing the light intensity around the photocells, which effectively decreased the resistive component of the photocells, the steady-state transmittance decreased as the energy of the incoming surface wave was strongly absorbed by the metasurface line (Figure S13b). These characteristics were also observed when both temperature-dependent capacitors and photocells were included within the metasurface line (Figure S13c). Thus, these simplified metasurface line samples also demonstrated that transmittance profiles can be controlled by introducing circuit components that behave differently in accordance with physical quantities such as temperature and light intensity.

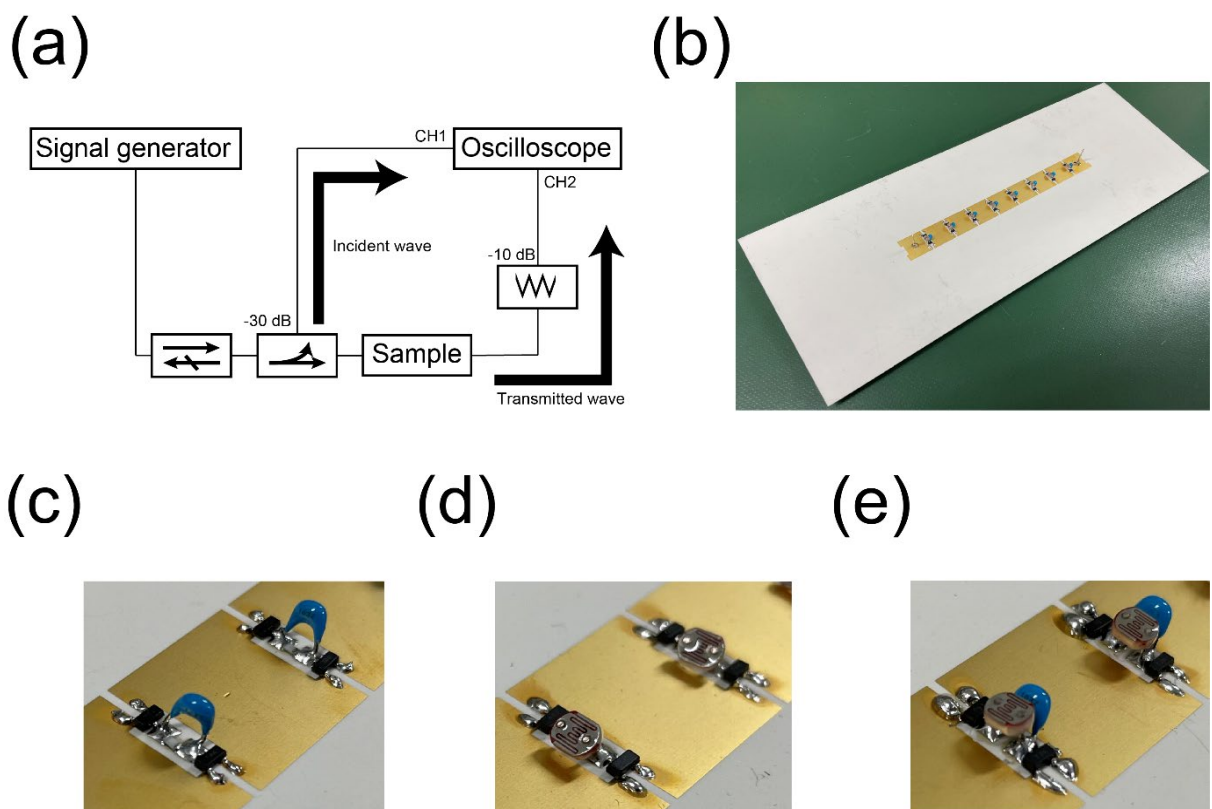

**Figure S12.** Measurement setup and samples for one-dimensional metasurfaces. (a) Measurement setup. (b) Example of the entire measurement sample. The samples were measured using (c) temperature-dependent capacitors and fixed resistors (10 kΩ), (f) fixed capacitors (1 nF) and photocells, and (d) both temperature-dependent capacitors and photocells.

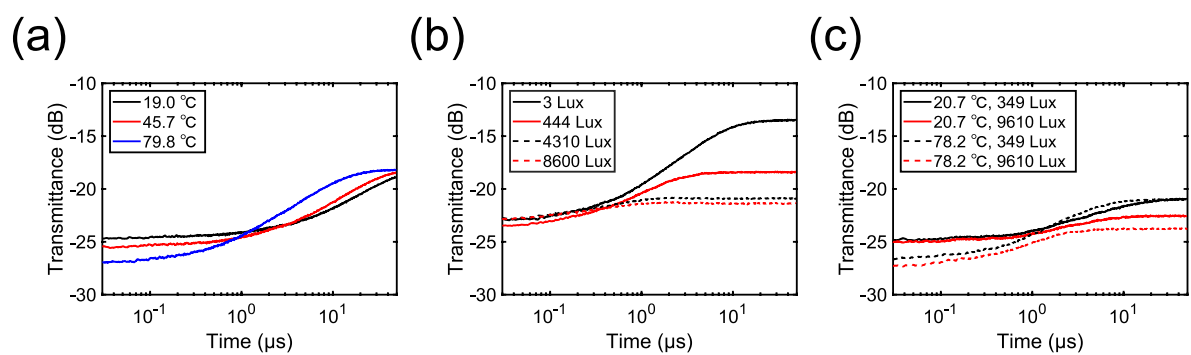

**Figure S13.** The measured transient transmittance results of the one-dimensional metasurface line samples shown in Figure S12. The results using (a) temperature-dependent capacitors and fixed resistors (10 kΩ), (b) fixed capacitors (1 nF) and photocells, and (c) both temperature-dependent capacitors and photocells at 4.81 GHz with 13 dBm. The measurement samples of (a) to (c) correspond to Figure S12c, Figure S12d, and Figure S12e, respectively.

**Supporting Note 4: Estimation of Physical Quantities**

This section provides additional information on the estimation of physical quantities. In Figure 4, we reported that temperature and light intensity were very well predicted, with good determination coefficients ranging from approximately 0.96 to 0.99. However, as seen in other machine learning studies,[46–48] the relationship between the number of training datasets $N_{tr}$ and that of test datasets $N_{te}$ is important for accurately estimating temperature and light intensity. For instance, Figure S14 and Figure S15 show the results for varying relationships between these two numbers for temperature and light intensity estimations, respectively, while Table S7 and Table S8 show the specific values used for $N_{tr}$ and $N_{te}$ in Figure S14 and Figure S15, respectively, as well as their determination coefficients. Here, the number of entire datasets, including both $N_{tr}$ and $N_{te}$, was fixed at 2290, and the training datasets were selected at random using a built-in function in Python (specifically, "split"), which was the same condition as that adopted in Figure 4. However, as the number of test datasets relatively increased, the determination coefficient decreased, as shown in Figure S14 and Figure S15 (also shown in Table S7 and Table S8). This relationship is more clearly shown in Figure S16, where the determination coefficient is plotted as a function of the number of training datasets. These results ensure that the training process needs to be conducted sufficiently so that a good determination coefficient is maintained. Essentially, a similar conclusion may be drawn via a different approach, as shown in Figure S17, Figure S18 and Figure S19, where temperature and light intensity were estimated using ridge regression instead of random forest. These determination coefficients and the number of test datasets used are summarized in Table S9 and Table S10. Compared with the results of random forest regression in Figure S16, the estimation results using ridge regression showed larger determination coefficients with a small number of training datasets. For instance, with only $N_{tr}$ = 11, the determination coefficients of ridge regression were greater than 0.9, while those of random forest regression remained at approximately 0.6. These results indicate that ridge regression is superior to random forest regression if the number of training datasets is limited, as ridge regression is suitable for avoiding overtraining and effectively building a model representing the relationship between the reflected waveform and physical parameters (i.e., temperature and light intensity). However, by increasing the number of training datasets to $N_{tr}$ = 2175, the determination coefficients of random forest regression exceeded those of ridge regression because random forest regression tends to derive a more accurate model by further increasing the training datasets. All these results indicate that machine learning-based estimation methods need to be properly adopted depending on the situation, such as the number of training datasets available and the estimation accuracy needed.

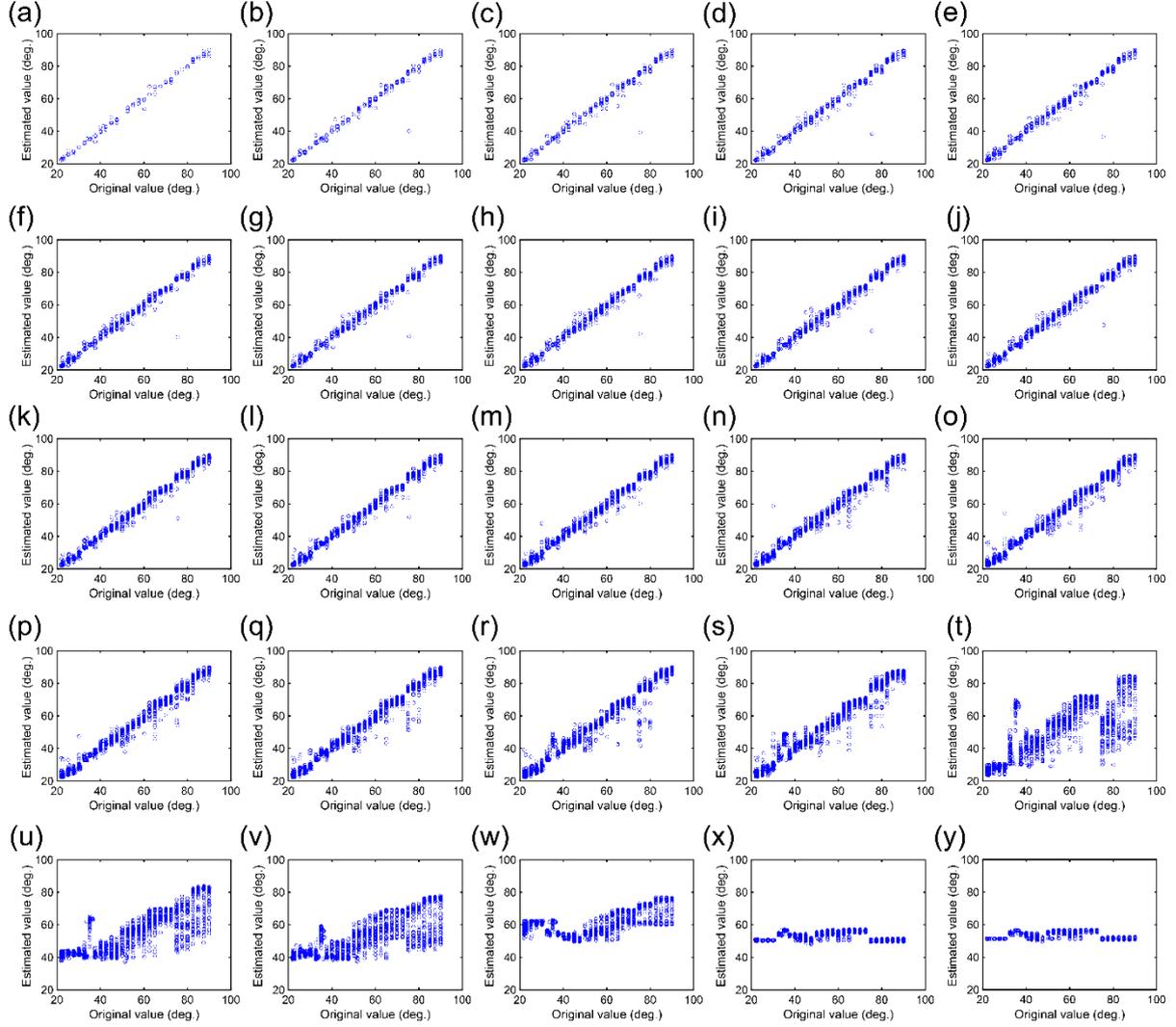

**Figure S14.** Estimation results of temperature in the random forest regression with different numbers of test datasets $N_{te}$ and training datasets $N_{tr}$. $N_{te}$ = (a) 115, (b) 229, (c) 344, (d) 458, (e) 573, (f) 687, (g) 802, (h) 916, (i) 1031, (j) 1145, (k) 1260, (l) 1374, (m) 1489, (n) 1603, (o) 1718, (p) 1832, (q) 1947, (r) 2061, (s) 2176, (t) 2268, (u) 2279, (v) 2281, (w) 2284, (x) 2286, and (y) 2288. The corresponding determination coefficients are shown in Table S7.

**Table S7.** The numbers of test datasets and training datasets used for Figure S14 (temperature estimation) and determination coefficients. The relationship between the number of training datasets and determination coefficients is shown in Figure S16.

| Panel | Number of test datasets $N_{te}$ | Number of training datasets $N_{tr}$ | Determination coefficient for temperature estimation |
|-------|-------|-------|-------|
| (a) | 115 | 2175 | 0.9958 |
| (b) | 229 | 2061 | 0.9828 |
| (c) | 344 | 1946 | 0.9853 |
| (d) | 458 | 1832 | 0.9865 |
| (e) | 573 | 1717 | 0.9882 |
| (f) | 687 | 1603 | 0.9898 |
| (g) | 802 | 1488 | 0.9905 |
| (h) | 916 | 1374 | 0.9910 |
| (i) | 1031 | 1259 | 0.9912 |
| (j) | 1145 | 1145 | 0.9924 |
| (k) | 1260 | 130 | 0.9927 |
| (l) | 1374 | 916 | 0.9921 |
| (m) | 1489 | 801 | 0.9919 |
| (n) | 1603 | 687 | 0.9881 |
| (o) | 1718 | 572 | 0.9870 |
| (p) | 1832 | 458 | 0.9861 |
| (q) | 1947 | 343 | 0.9757 |
| (r) | 2061 | 229 | 0.9627 |
| (s) | 2176 | 114 | 0.9622 |
| (t) | 2268 | 22 | 0.6047 |
| (u) | 2279 | 11 | 0.6456 |
| (v) | 2281 | 9 | 0.5492 |
| (w) | 2284 | 6 | 0.2237 |
| (x) | 2286 | 4 | -0.0350 |
| (y) | 2288 | 2 | -0.0203 |

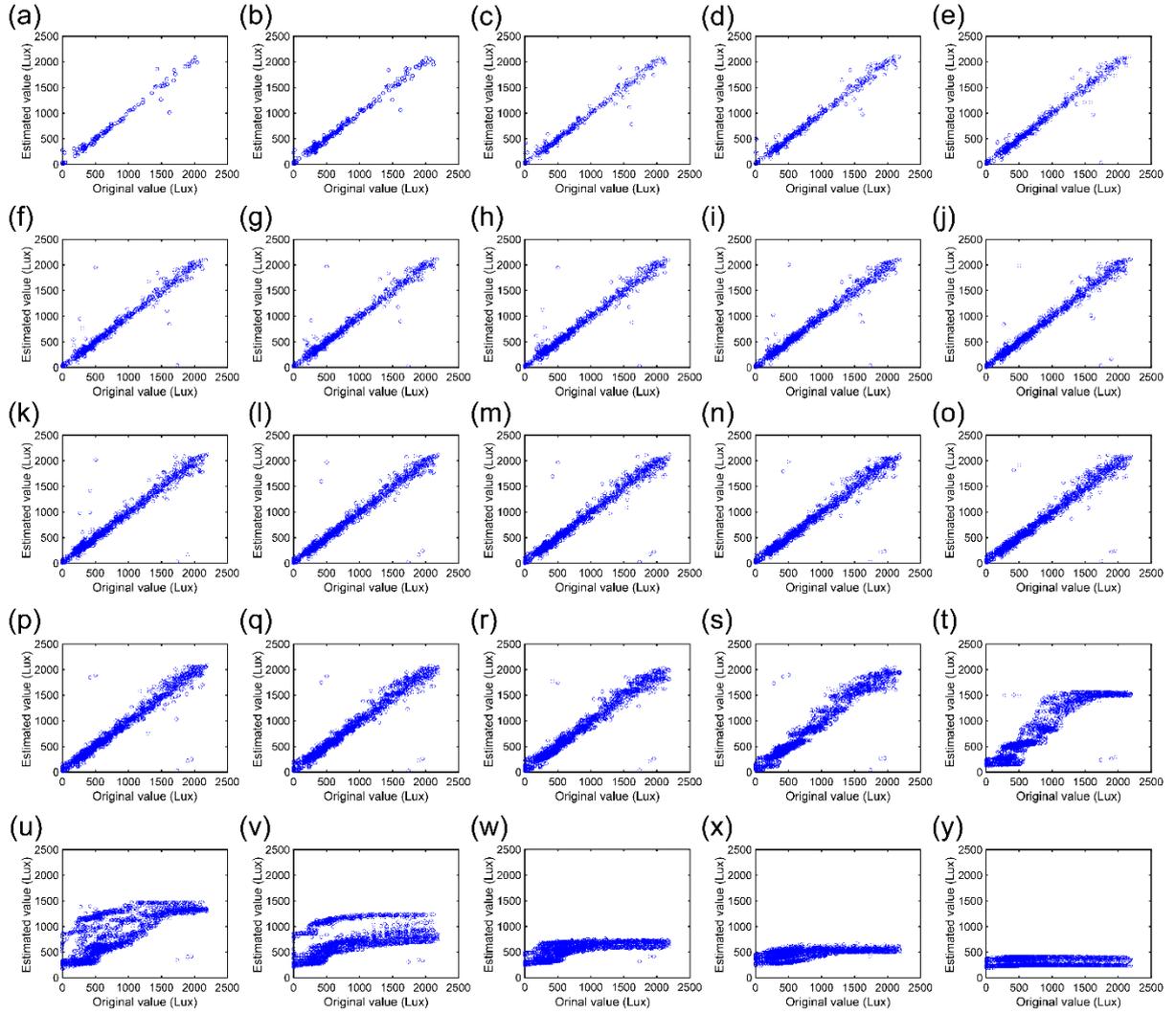

**Figure S15.** Estimation results of light intensity in the random forest regression with different numbers of test datasets $N_{te}$ and training datasets $N_{tr}$. $N_{te}$ = (a) 115, (b) 229, (c) 344, (d) 458, (e) 573, (f) 687, (g) 802, (h) 916, (i) 1031, (j) 1145, (k) 1260, (l) 1374, (m) 1489, (n) 1603, (o) 1718, (p) 1832, (q) 1947, (r) 2061, (s) 2176, (t) 2268, (u) 2279, (v) 2281, (w) 2284, (x) 2286, and (y) 2288. The corresponding determination coefficients are shown in Table S8.

**Table S8.** The numbers of test datasets and training datasets used for Figure S15 (light-intensity estimation) and determination coefficients. The relationship between the number of training datasets and determination coefficients is shown in Figure S16.

| Panel | Number of test datasets $N_{te}$ | Number of training datasets $N_{tr}$ | Determination coefficient for light-intensity estimation |
|---|---|---|---|
| (a) | 115 | 2175 | 0.9714 |
| (b) | 229 | 2061 | 0.9831 |
| (c) | 344 | 1946 | 0.9774 |
| (d) | 458 | 1832 | 0.9800 |
| (e) | 573 | 1717 | 0.9677 |
| (f) | 687 | 1603 | 0.9583 |
| (g) | 802 | 1488 | 0.9633 |
| (h) | 916 | 1374 | 0.9643 |
| (i) | 1031 | 1259 | 0.9678 |
| (j) | 1145 | 1145 | 0.9637 |
| (k) | 1260 | 130 | 0.9627 |
| (l) | 1374 | 916 | 0.9601 |
| (m) | 1489 | 801 | 0.9597 |
| (n) | 1603 | 687 | 0.9607 |
| (o) | 1718 | 572 | 0.9611 |
| (p) | 1832 | 458 | 0.9610 |
| (q) | 1947 | 343 | 0.9612 |
| (r) | 2061 | 229 | 0.9566 |
| (s) | 2176 | 114 | 0.9434 |
| (t) | 2268 | 22 | 0.8665 |
| (u) | 2279 | 11 | 0.5841 |
| (v) | 2281 | 9 | 0.2252 |
| (w) | 2284 | 6 | 0.0907 |
| (x) | 2286 | 4 | -0.1625 |
| (y) | 2288 | 2 | -0.7480 |

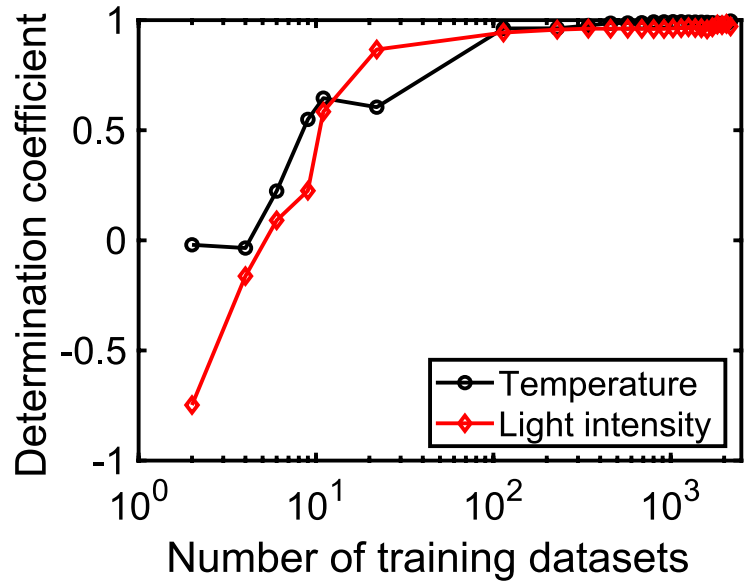

**Figure S16.** Determination coefficients of random forest regression for different numbers of training datasets. The relationships between original values and estimated values are shown in Figure S14 and Figure S15 for temperature and light-intensity estimations, respectively.

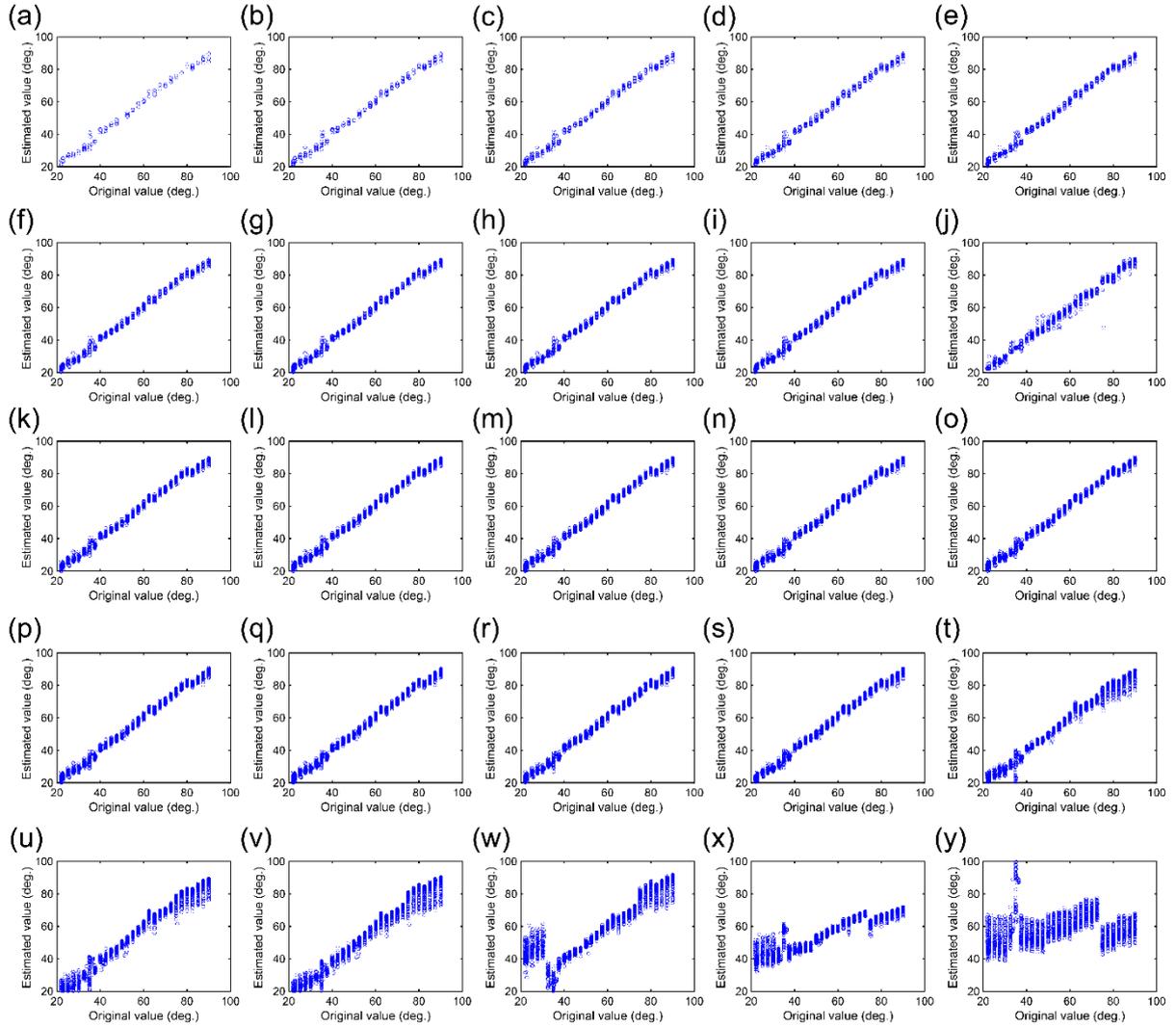

**Figure S17.** Estimation results of temperature in the ridge regression with different numbers of test datasets $N_{te}$ and training datasets $N_{tr}$. $N_{te}$ = (a) 115, (b) 229, (c) 344, (d) 458, (e) 573, (f) 687, (g) 802, (h) 916, (i) 1031, (j) 1145, (k) 1260, (l) 1374, (m) 1489, (n) 1603, (o) 1718, (p) 1832, (q) 1947, (r) 2061, (s) 2176, (t) 2268, (u) 2279, (v) 2281, (w) 2284, (x) 2286, and (y) 2288. The corresponding determination coefficients are shown in Table S9.

**Table S9.** The numbers of test datasets and training datasets used for Figure S17 (temperature estimation) and determination coefficients. The relationship between the number of training datasets and determination coefficients is shown in Figure S19.

| Panel | Number of test datasets $N_{te}$ | Number of training datasets $N_{tr}$ | Determination coefficient for temperature estimation |
|---|---|---|---|
| (a) | 115 | 2175 | 0.9933 |
| (b) | 229 | 2061 | 0.9934 |
| (c) | 344 | 1946 | 0.9938 |
| (d) | 458 | 1832 | 0.9936 |
| (e) | 573 | 1717 | 0.9941 |
| (f) | 687 | 1603 | 0.9940 |
| (g) | 802 | 1488 | 0.9942 |
| (h) | 916 | 1374 | 0.9942 |
| (i) | 1031 | 1259 | 0.9939 |
| (j) | 1145 | 1145 | 0.9924 |
| (k) | 1260 | 130 | 0.9939 |
| (l) | 1374 | 916 | 0.9939 |
| (m) | 1489 | 801 | 0.9939 |
| (n) | 1603 | 687 | 0.9939 |
| (o) | 1718 | 572 | 0.9938 |
| (p) | 1832 | 458 | 0.9938 |
| (q) | 1947 | 343 | 0.9938 |
| (r) | 2061 | 229 | 0.9936 |
| (s) | 2176 | 114 | 0.9935 |
| (t) | 2268 | 22 | 0.9792 |
| (u) | 2279 | 11 | 0.9691 |
| (v) | 2281 | 9 | 0.9615 |
| (w) | 2284 | 6 | 0.8003 |
| (x) | 2286 | 4 | 0.6731 |
| (y) | 2288 | 2 | -0.0487 |

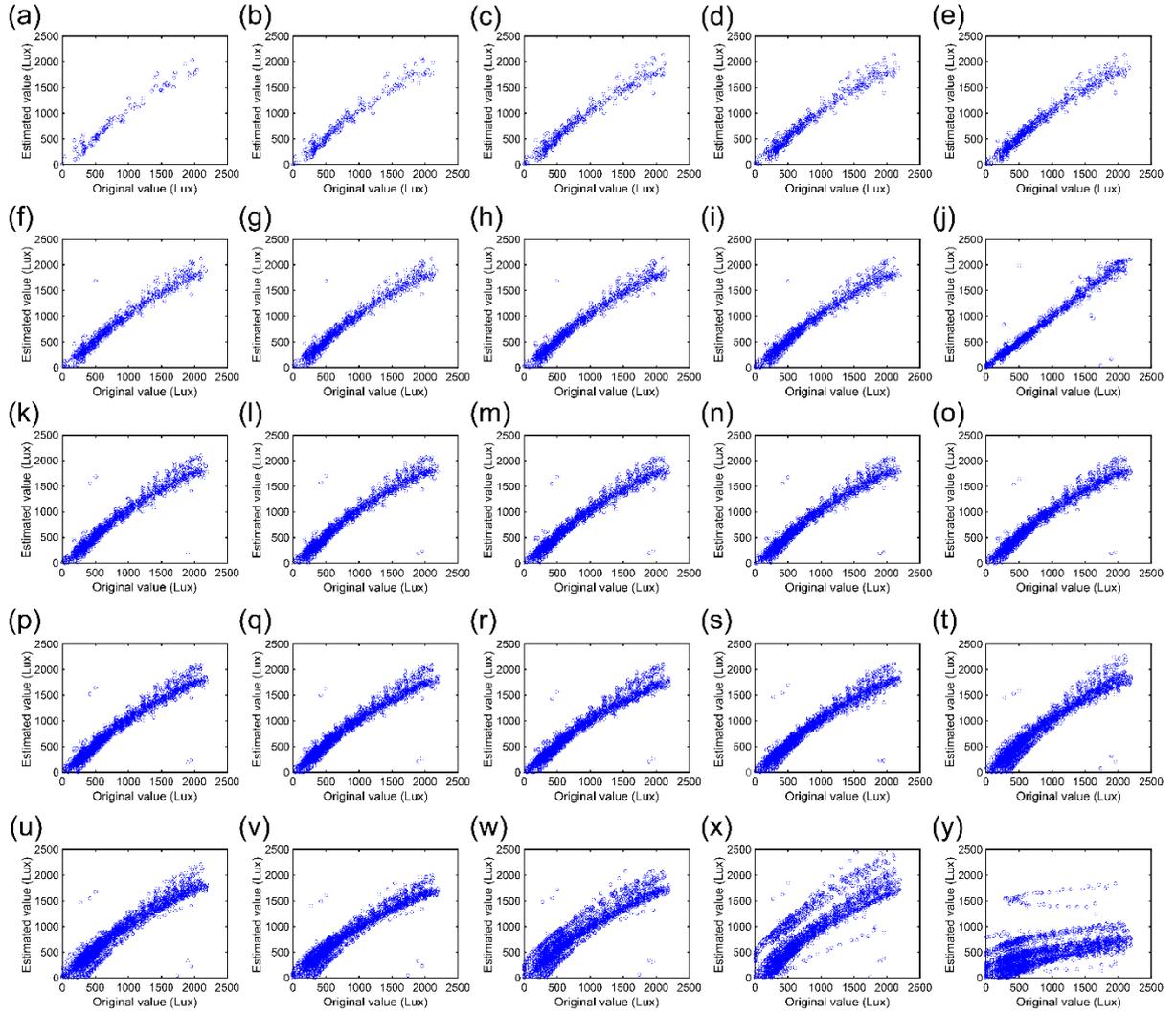

**Figure S18.** Estimation results of light intensity in the ridge regression with different numbers of test datasets $N_{te}$ and training datasets $N_{tr}$. $N_{te}$ = (a) 115, (b) 229, (c) 344, (d) 458, (e) 573, (f) 687, (g) 802, (h) 916, (i) 1031, (j) 1145, (k) 1260, (l) 1374, (m) 1489, (n) 1603, (o) 1718, (p) 1832, (q) 1947, (r) 2061, (s) 2176, (t) 2268, (u) 2279, (v) 2281, (w) 2284, (x) 2286, and (y) 2288. The corresponding determination coefficients are shown in Table S10.

**Table S10.** The numbers of test datasets and training datasets used for Figure S18 (light-intensity estimation) and determination coefficients. The relationship between the number of training datasets and determination coefficients is shown in Figure S19.

| Panel | Number of test datasets $N_{te}$ | Number of training datasets $N_{tr}$ | Determination coefficient for light-intensity estimation |
|---|---|---|---|
| (a) | 115 | 2175 | 0.9556 |
| (b) | 229 | 2061 | 0.9582 |
| (c) | 344 | 1946 | 0.9580 |
| (d) | 458 | 1832 | 0.9604 |
| (e) | 573 | 1717 | 0.9436 |
| (f) | 687 | 1603 | 0.9411 |
| (g) | 802 | 1488 | 0.9444 |
| (h) | 916 | 1374 | 0.9472 |
| (i) | 1031 | 1259 | 0.9487 |
| (j) | 1145 | 1145 | 0.9637 |
| (k) | 1260 | 130 | 0.9416 |
| (l) | 1374 | 916 | 0.9360 |
| (m) | 1489 | 801 | 0.9372 |
| (n) | 1603 | 687 | 0.9386 |
| (o) | 1718 | 572 | 0.9390 |
| (p) | 1832 | 458 | 0.9405 |
| (q) | 1947 | 343 | 0.9406 |
| (r) | 2061 | 229 | 0.9655 |
| (s) | 2176 | 114 | 0.9374 |
| (t) | 2268 | 22 | 0.9227 |
| (u) | 2279 | 11 | 0.9148 |
| (v) | 2281 | 9 | 0.9070 |
| (w) | 2284 | 6 | 0.8744 |
| (x) | 2286 | 4 | 0.7500 |
| (y) | 2288 | 2 | -0.0619 |

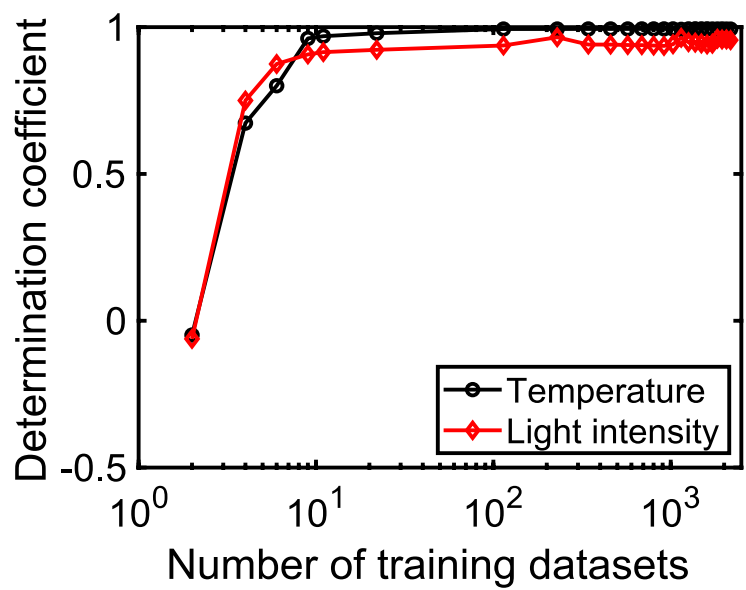

**Figure S19.** Determination coefficients of ridge regression for different numbers of training datasets. The relationships between original values and estimated values are shown in Figure S17 and Figure S18 for temperature and light-intensity estimations, respectively.